\documentclass[a4, 11pt]{article}

\usepackage{graphicx}

\usepackage{amsmath}

\usepackage{caption}
\usepackage{cochineal} 
\usepackage[protrusion=true]{microtype} 
\usepackage{geometry} 
\usepackage{hyperref}
 
\usepackage[hang,flushmargin]{footmisc} 

\usepackage[greek,english]{babel}
\usepackage{csquotes}

\usepackage{gensymb}

\usepackage[style=windycity, style=authoryear]{biblatex}
\begin{document}

\title{The logic of planetary combination\\in Vettius Valens }

\author{Claire Hall$^a$ and Liam P. Shaw$^b$}

\date{%
    \begin{small}
 
    $^a$ All Souls College, Oxford OX1 4AL\\%
    $^b$ Department of Zoology, University of Oxford, Oxford OX1 3SZ\\[2ex]%
    \end{small}
}

\maketitle

\begin{abstract}
\noindent The \textit{Anthologies} of the second-century astrologer Vettius Valens (120-c.175 CE) is the most extensive surviving practical astrological text from the period. Despite this, the theoretical underpinnings of the \textit{Anthologies} have been understudied; in general, the work has been overshadowed by Ptolemy's contemporaneous \textit{Tetrabiblos}. While the  \textit{Tetrabiblos} explicitly aims to present a systematic account of astrology, Valens' work is often characterised as a miscellaneous collection, of interest to historians only for the evidence it preserves about the practical methods used in casting horoscopes. In this article, we argue that the \textit{Anthologies} is also an invaluable resource for engagement with the conceptual basis of astrology. As a case study, we take a section of \textit{Anthologies} Book 1 which lists the possible astrological effects of planets, both alone and in `combinations' of two and three. We demonstrate that analysing Valens' descriptions quantitatively with textual analysis reveals a consistent internal logic of planetary combination. By classifying descriptive terms as positive or negative, we show that the resulting `sentiment' of planetary combinations is well-correlated with their component parts. Furthermore, we find that the sentiment of three-planet combinations is more strongly correlated with the average sentiment of their three possible component pairs than with the average sentiment of individual planets, suggesting an iterative combinatorial logic. Recognition of this feature of astrological practice has been neglected compared to the mathematical methods for calculating horoscopes. We argue that this analysis not only provides evidence that the astrological lore detailed in Valens is more consistent than is often assumed, but is also indicative of a wider methodological technique in practical astrology: combinatorial reasoning from existing astrological lore.\footnote{We thank Dmitri Levitin, Ed Godfrey, and two anonymous reviewers for helpful comments. Since quantitative analysis of this sort does not fit neatly into the usual format of a historical article, we have supplemented our article with code available at \href{https://github.com/liampshaw/vettius-valens}{https://github.com/liampshaw/vettius-valens}.}

\end{abstract}

\textbf{Correspondence:} claire.hall@all-souls.ox.ac.uk, liam.shaw@zoo.ox.ac.uk

\newpage

\section{Introduction}

The \textit{Anthologies} of Vettius Valens is the most extensive practical astrological text which survives from Greek antiquity.\footnote{In this article, references to the Greek text of the \textit{Anthologies} refer to the Teubner edition by \textcite{vettius_valens_antiochenus_vettii_2010} and English translations come from \textcite{riley_anthologies}. References to the Greek text of Ptolemy's \textit{Tetrabiblos} are to the Teubner edition by \textcite{ptolemy_apotelesmatica_hubner}, and English translations are unless otherwise stated from the Loeb edition of \textcite{robbins_tetrabiblos}.} The \textit{Anthologies} consists of nine books, which include a huge quantity of information on astrological methods as well as catalogues of predictions. In total, the \textit{Anthologies} contains around 130 partial or complete horoscopes ranging from 37-188 CE, some of which are used multiple times.\footnote{The number of unique horoscopes is 123. See \textcite[127]{boxer_scheme_2020} and \textcite[176]{neugebauer_greek_1959}.} These horoscopes suggest that the majority of the work was written between 152 and 162 CE, and the \textit{Anthologies} has been in almost constant use since.\footnote{For the date of the \textit{Anthologies}, see \textcite{neugebauer_chronology_1954}. For an extensive survey of the contents and composition of the \textit{Anthologies}, see \textcite{riley_survey_1996}. Neugebauer complained that the text `is in an abominable condition…the [numerical] examples are full of mistakes' \autocite[794]{neugebauer_history_1975}; for the manuscript history, see \textcite{pingree_byzantine_1983}.}  As a resource for `literary' horoscopes (i.e. those not from an original text, such as papyri) it is unparalleled: Neugebauer and Van Hoesen noted there would only be five other extant literary horoscopes dating from before 380 CE if the \textit{Anthologies} had not survived. \footnote{\textcite[176]{neugebauer_chronology_1954}. For this reason, the \textit{Anthologies} represented `a rich quarry of new or rare words' for lexicographers when Kroll's first edition was published \autocite[6]{jones_making_1941}. \textcite{greenbaum_hellenistic_2020} gives an introduction to wider Hellenistic horoscopes. See also the authoritative catalogue of horoscopes in \textcite[213-316]{HeilenStephan2015} and the analysis at pp. 523-4.} Although we have little external biographical information about Vettius Valens, one of these horoscopes was suggested by Pingree to be that of Valens himself.\footnote{See \textcite[v]{vettius_valens_antiochenus_vettii_2010}. This claim is now widely accepted, see e.g. \textcite{greenbaum_vettius_2019}.} From this horoscope and the associated details Valens provides, we can infer that he was born on 8 February 120 CE in Antioch, travelled widely, including to Egypt, and spent the majority of his working life as an astrologer in Alexandria.\footnote{Valens refers regularly to horoscopes that he himself has cast, including at \textit{Anthologies} 7.3.255; 7.3.256; 7.6.271.}

Valens is often contrasted with his near-contemporary Ptolemy (c.100–c.170 CE), also associated with Alexandria. Although both wrote on astrology, their styles are markedly different. In particular, Valens' work is often characterised as practical, in opposition to the more theoretical work of Ptolemy's \textit{Tetrabiblos}. According to Neugebauer, the \textit{Anthologies} and the \textit{Tetrabiblos} were written almost simultaneously but exhibit a strong difference in style and purpose: Valens' work seeks to `confirm and refine growing [astrological] doctrine', whereas Ptolemy's seeks to form a `consistent theory of… a universal science of life.' \autocite[67]{neugebauer_chronology_1954}. Riley makes a similar case, characterising Ptolemy's work as `axiomatic' in its method: Ptolemy `lays down principles that are universally true, then he applies them in individual cases'; whereas Valens is `quasi-exhaustive', constructing `listings of configurations' that go on `at great length' \autocite[249]{riley_theoretical_1987}. 

Given this contrast, if Ptolemy is a systematic philosophical thinker it may be tempting to treat Valens as his opposite: a miscellanist, churning out regurgitated screeds of astrological lore. This picture may also be influenced by our knowledge of Ptolemy's wide variety of other interests, primarily in astronomy but also in geography, harmonics, and optics (the latter two in lost works).\footnote{For an introduction to Ptolemy's life and context, see \textcite{feke_ptolemys_2019}. For more on the wider context of Ptolemy's astronomical theory, see \textcite{taub_ptolemys_1993}. See also \textcite{JonesAlexander2010}.} Indeed, even within the \textit{Tetrabiblos} Ptolemy considers wide areas of astrological inquiry which are absent from the \textit{Anthologies}, devoting large sections to `general' astrology, which pertains to predictions that affect `whole races, countries, and cities'\footnote{Ptolemy, \textit{Tetrabiblos} 2.1.2 H88, R119.} and including geographical material and predictions of the weather. By contrast, the \textit{Anthologies} is presented by Valens himself as a handbook for practical astrology. 

Scholars have not been impressed by Valens as a thinker: \textcite{cumont_egypt_1937} called him `narrow-minded and devoid of originality' (p. 18)\footnote{`\textit{esprit borné, dépourvu de tout originalité}', quoted in \textcite[5]{riley_survey_1996}.} and \textcite{riley_theoretical_1987} notes that when Valens attempts `empirical' proofs of his craft, these are little more than the retrofitting of observed events to his own previously cast horoscopes (p. 248).\footnote{Riley's assessments elsewhere are more generous, noting that while he was `no creative genius', his energy `cannot be faulted' \autocite{riley_survey_1996}. \textcite{neugebauer_history_1975} also appears to have been unimpressed by Valens as an author, commenting that `[o]ne may doubt whether Vettius Valens was aware of such inconsistencies in different doctrines [on rising times] he had inherited from his predecessors' (p. 719), that Valens `apparently did not understand that [certain parameters] were underlying his rules' (p. 794), and that he has `a certain tendency to present supposedly ``handy'' procedures in an unnecessarily devious fashion' (p. 823). This dim view of Valens has been shared by many scholars. An early review of Kroll's edition notes that `[Valens] was not equal to the task of digesting his sources and arranging his material satisfactorily in chapters as he wished to do' \autocite[526]{kroll_review_1910}, and \textcite[455]{winlock_origin_1940} suggests that Valens' understanding of Egyptian lunar months was based on `some misunderstanding\ldots[of a] popular saying'.} It is certainly true that Valens does not attempt—or seem to show any interest in—either a detailed philosophical account of the physics of the heavens or a wider investigation of the implications of astrology. This broad contrast between practical and theoretical often carries a (sometimes explicit) value judgement in Ptolemy's favour \autocite[236]{riley_theoretical_1987}.  Nevertheless, it does not follow from this contrast in method and aims that Valens' work was free of conceptual underpinnings. Nor does it follow that it was unsystematic or unscientific – a suggestion implicit in Riley and others.\footnote{A comparison can be made with the Greek medical tradition, in which some writers (Serapion of Alexandria, Sextus Empiricus) became known as the Empirics, taking an approach which declined to speculate{---}beyond certain bounds{---}on matters such as the underlying physical theories of disease-causation; in contrast, other writers later termed as `dogmatists' (Hippocratics, Herophilus, Erasistratus) argued that medicine should aim not just at the practical craft of healing, but at a total theoretical and physical account of nature with respect to sickness and health. Something similar does seem to be at play among astrological writers, with Ptolemy's method and aims mirroring those of the dogmatists, while most other astrological writers (Valens, Dorotheus of Sidon, Firmicus Maternus) taking a much more empiricist approach. For Greek medicine more generally, see \textcite{nutton_ancient_2004}. For medical empiricism, see \textcite{frede_ancient_1987}.} In this article we make the case that Valens' lists of planetary combinations show a greater degree of cohesion than has been previously noted. We analyse the descriptions quantitatively to show that they are not merely miscellaneous assortments of associations, but exhibit a strong internal logic. Despite the divergent methods and aims on display on the textual surface of Ptolemy and Valens, Valens' general understanding of the underlying structure of astrology may be closer to Ptolemy than usually assumed. 

\subsection{Quantitative methods in the history of astrology}

The modern mathematical study of Greek astrology owes much to the twentieth century historian Otto Neugebauer, who trained as a mathematician.\footnote{Modern historical study of Greek astrology had been undertaken from the 19\textsuperscript{th} century onwards. See \textcite[3-9]{HeilenStephan2015}.} While Neugebauer acknowledged the value of astrology for social history{--}its insight into `the daily life, religion and superstition... and cosmogonic ideas' of the past \autocite{neugebauer_study_1951}{--}he emphasised that the primary justification for studying it was because of its value for the history of astronomy and the exact sciences. He argued that astrological texts offer a window onto the mathematical knowledge of their period: `[t]he only hope of obtaining a few glimpses of the astronomical methods of the time of Hipparchus rests in the painstaking investigation of wretched writers like Vettius Valens' \autocite{neugebauer_study_1951}.\footnote{He was using the term `wretched' with irony: his defence of `wretched subjects' came in response to George Sarton and Frances Siegel's summary of E. S. Drower's edition of the Mandean \textit{Book of the Zodiac} which described it as `a wretched collection of omens, debased astrology, and miscellaneous nonsense' \textcite[374]{sarton_seventy-sixth_1950}. Neugebauer accepted this description as `factually correct', but not one telling the whole story \textcite{neugebauer_study_1951}. Although such `wretched' texts were viewed as a wrong turn in the history of science, they remained beyond the mathematical comprehension of many historians who held this opinion. In this respect Neugebauer was a notable exception, although ignorance about the astronomical basis of astrological texts persisted.} As late as 1942, Neugebauer could observe in passing that `one of the most important sources on Hellenistic-Oriental astronomy, Vettius Valens\ldots is practically unexamined, as far as information about exact astronomy is concerned. This problem very much deserves serious consideration' \autocite[239]{neugebauer_egyptian_1942}. 

Until the mid-twentieth century it was unknown whether the horoscopes in the \textit{Anthologies} referred to real and contemporary planetary configurations. In a pioneering work, \textcite{neugebauer_greek_1959} showed how to ascertain the exact date of even partial literary horoscopes. Their method was manual and required aligning the motions of planets on long sheets of graph paper.\footnote{The method is worth recounting to appreciate what was involved in undertaking of this work without computational methods. It involved taking a `long sheet of graph paper...divided into 60 units of length, representing 60 years, and into 12 units of width, representing the twelve zodiacal signs' \autocite[1]{neugebauer_greek_1959}. Then, they prepared two graphs of the mean motion of Saturn and Jupiter on tracing paper using the same scale, and overlaid them on this sheet of graph paper after calibrating their end positions to known dates. These computations for planets were carried out using tables of positions accurate to about one degree, from published astronomical tables. Once the graphs had been overlaid, this allowed them to identify possible dates for a horoscope within the 60-year period: for the simplest case, a horoscope with a conjunction of Saturn and Jupiter meant that only dates where the graphs of Saturn and Jupiter intersected were possible. They could therefore `home in' on possible dates, using other considerations to rule some dates out (e.g. dates later than the composition of the \textit{Anthologies}). Their analysis was simplified as it progressed by the observation that each chapter of the \textit{Anthologies} had narrow limits of dates, suggesting composition at separate times.} In most cases this permitted them to identify a precise date to an accuracy of around a day. Overall, this successfully led to the identification of dates for almost all horoscopes in the \textit{Anthologies}, and put the empirical basis of Valens' work beyond question. Although it would have been easy to invent horoscopes for pedagogical purposes, Valens did not do so. 

The modern availability of accurate ephemerides such as the Swiss Ephemeris \autocite{koch_swiss_nodate} together with computational libraries for manipulating horoscopes means that reproducing Neugebauer and Van Hoesen's work is relatively straightforward. However, our focus here is not the dating of individual horoscopes, but the use of quantitative techniques for aiding textual analysis. Lynn Thorndike applied a similar approach to Firmicus Maternus, manually counting occurrences of terms to introduce a `quantitative element': `one naturally assumes that those matters to which Firmicus devotes most space and emphasis are the most prominent features of his age' \autocite[417]{thorndike_roman_1913}. We see this as a precursor to the analysis we present here.\footnote{Valens was used as a source for social history by \textcite{macmullen_social_1971}. However, \textcite{riley_survey_1996} cautions against the naive view that astrologers are simply `reflectors of their own society' (p. 12). Here, we are analysing Valens as a source for astrology and not for wider society, and we recognise that different parts of Valens' doctrine derive from sources and traditions of different ages.} 

We are not suggesting that this quantitative analysis reveals details which cannot be discovered by other methods, nor that it is more sophisticated: it is a blunter instrument than close reading. But quantitative textual analysis is a historical method well-suited to uncovering the broad features of a body of text, and the inter-relationships between its sections, which may not be apparent from a linear reading.\footnote{For the history of this kind of quantitative reading, see \textcite[14-37]{DobsonJamesE2019CDHT}. See also \textcite[21-43]{crymbletechnology}.} This is particularly true for a writer like Valens: it is difficult to know whether an unfocused style may be obscuring a greater underlying cohesion in content than is apparent. We would like to suggest that these methods have wider applicability to the investigation of astrological texts; however, as an initial case study we apply them narrowly to a situation where one would expect to uncover relationships: Valens' descriptions of the qualities of the planets and of their combinations.

\subsection{The planets and their qualities}

The hierarchy of planets in astrology seems to have been established very early in its history and persisted:

\begin{quote}
    \centering Saturn, Jupiter, Mars, Venus, Mercury.
\end{quote}

This order is found across astrological texts from Babylon, Egypt and Greece\footnote{See \textcite[164]{neugebauer_greek_1959} for a discussion of planetary order in papyri and literary Greek horoscopes. They note that the Babylonian horoscopes have a different standard order of enumeration: Jupiter, Venus, Mercury, Saturn, Mars. Demotic horoscopes enumerate centers and related places in a fixed order, mentioning planets when they coincide with a place.} and continued to be accepted.\footnote{Over two centuries after Valens, Augustine rhetorically asked why Jupiter{---}the king of the gods{---}should be less bright than Venus and less high up in the sky than Saturn (\textit{City of God} 7.15 \autocite{augustine_city_II}; by `less high up in the sky', he means far away from earth). His aim was polemical. As a former Manichee who probably trained in astrology (see \textcite{van_oort_augustines_2011}), Augustine knew that the planets were conventionally listed in the above order.} The reason for this consistency may have pragmatic origins: it matches their distances from the Sun (in a heliocentric model), from furthest to nearest, and so corresponds to their speed of movement over the year through the zodiac.\footnote{This order is so prevalent it has been referred to as the `customary cosmological order' e.g. \textcite{freeth_model_2021}. It should be noted that while the superior planets (Mars, Jupiter, Saturn) have progressively slower periods, matching this order, the inferior planets (Mercury and Venus) take on average the same time to travel through the zodiac as the Sun.}  This order of the planets is largely consistent.\footnote{So much so, that it can be used to identify copying errors in manuscripts when planets appear out of order (e.g. discussing the dating of a particular horoscope in the \textit{Anthologies}, \textcite[66]{neugebauer_chronology_1954} notes: `This also explains why Venus is mentioned out of order before Mars and right after Jupiter').}

In classical astrology, `planets' also included the two luminaries, the Sun and Moon, making seven classical planets. The order including the luminaries is more variable. While Valens uses the standard ordering of the five non-luminary planets, he tends to place the luminaries at the start when describing a horoscope,\footnote{That is: the Sun, the Moon, Saturn, Jupiter, Mars, the Sun, Venus, Mercury.} proceeding through the planets in this order but mentioning other planets when they occupy the same sign of the zodiac. In this article, when discussing configurations of planets we use the more normal astrological ordering as accepted by Ptolemy\footnote{Ptolemy took this as his actual cosmological model in terms of distance from the Earth, arguing `that Mercury and Venus lie between the moon and the sun, even though\ldots [as he admitted] no observations had yet been made of their occultation of the sun', based on a combination of aesthetic and theoretical grounds \autocite[193]{feke_ptolemys_2019}. It should also be noted that sometimes in Hellenistic horoscopes when the planets are listed with the two luminaries, the luminaries can be placed at the end of the list.} 

\begin{quote}
    \centering Saturn, Jupiter, Mars, the Sun, Venus, Mercury, the Moon
\end{quote}

The order of the five non-luminary planets is consistent. But what of their qualities? Because of the continuous popularity of astrology from ancient times to our own, many of the characteristics of the planets in Greek astrology remain familiar to us.  The idea of Mars as war-like, or Saturn as cold and elderly, were, in Greek thought, deeply mythically rooted. Indeed, Riley notes that similarities in the mythic characteristics ascribed to planets and signs in Valens are also accepted by Ptolemy, suggesting their ubiquity \autocite[68]{riley_science_1988}. Such widespread acceptance coupled with a modern familiarity has probably occluded further study of planetary characteristics.\footnote{There is no comprehensive study, unlike that of H\"{u}bner for the signs of the zodiac, \autocite{HübnerWolfgang1982DEdT}.} Yet it has been established that alongside the basic characterisations, Babylonian, Egyptian, and Greek sources give a varying series of associations and characteristics for the planets. Some of these are repeated and popular; some appear only rarely. Commenting on this tendency in the case of Saturn, Bouché-Leclercq wrote that, from such a vast store of examples, `astrologers chose according to their liking how to compose the type of this powerful and dreaded star'.\footnote{`\textit{Astrologues ont choisi à leur gré de quoi composer le type de l'astre puissant et redouté}' \autocite{bouche-leclercq_astrologie_2014}.}
 
Partly due to these sorts of mythic characterisations, the view of the qualitative aspects of astrology as whimsical or pedantic often still prevails. There has been little systematic effort to investigate the internal conceptual consistency of Greek astrological interpretation – even of particular astrologers, with the exception of Ptolemy. Any focus on consistency in the scholarly literature is usually on the mathematical or theoretical aspects. This is not surprising, and is likely to be the result of two main trends. First, the interpretative claims of astrology still carry weight, and, whether implicitly or explicitly, most scholars seek to distance themselves from the central claim that the stars can cause or signify human events.\footnote{There are of course, difficult boundaries to contend with: most of us would accept that the tides and some weather phenomena are caused by the moon and sun.} Second, even with Greco-Roman divinatory methods that are more straightforwardly of the past, study of the precise claims of predictive methods is grounded by a scepticism which discourages taking particular interpretative claims too seriously – if the claims are false, then why suppose that they must be conceptually cohesive? The Ptolemaic approach placing planets within the scheme of Aristotelian physics is the exception, because of the clear link to wider Greek science. To modern readers it often appears a refreshing contrast to arbitrary lists of associations due to its physical basis. 

However, this distinction must not be overstated. Overall, there is a complex relationship between the general mythic characteristics of the planets, more narrowly-defined astrological traditions about them, and the physical claims made by astrologers about the planets' natures and characteristics.\footnote{We do not investigate mythic qualities here. However, one can imagine that where two planets (or their corresponding gods) have particularly strong mythic associations in combination, these might be expected to be reflected in astrological predictions. For example, one might expect the myth of the adultery of Venus and Mars to mean that adultery features in predictions from the astrological conjunction of Venus and Mars; alternatively, one might expect something to do with usurpation or parricide from conjunctions of Jupiter and Saturn on the same grounds. Neither is present in Valens. However, we have not conducted a thorough survey of all such associations. (With many thanks to the anonymous reviewer for this point).} Riley recognises the first and third of these categories, noting that sometimes in Ptolemy's \textit{Tetrabiblos} there is a slightly uneasy relationship between the two, and maintaining that Ptolemy distinguished between them, even if not in the terminology of `mythical' vs `physical' \autocite[68]{riley_science_1988}.  For Ptolemy, the physical characteristics of the planets fit into an Aristotelian system of balanced opposites: characteristics included physical qualities like hot and cold, wet and dry, as well as biological characteristics like male and female, and temporal characteristics like diurnal and nocturnal.\footnote{\textit{Tetrabiblos} 1.4 H22; 1.6-7 H27-29. For Aristotle's physical theory and how it relates to his element theory, see \textcite{lewis_aristotles_2018}.} The physical characteristics of the planets have an effect on the Earth: `Venus adds heat to the prevailing conditions, Saturn cold, Jupiter moisture, Mars dryness, Mercury motion and wind.'\footnote{\textit{Phases} 8. H11.24-27. Translation from \textcite[243]{riley_ptolemys_use}.}

Ptolemy also designates some secondary characteristics which are easy to categorise, including each planet's status as broadly beneficent or maleficent. It is never explicitly stated whether this primarily means beneficent towards human beings; Ptolemy does seem to have in mind also the relationship of the planets to one another. In Ptolemy's system the benefics are Jupiter, Venus, and the moon; the malefics are Saturn and Mars; the neutrals are the Sun and Mercury. While Ptolemy attributes these categorisations to `the ancients' (\textgreek{οἱ παλαιοί})  he also gives an Aristotelian physical explanation: those planets which have predominantly cold or dry natures are malefic.\footnote{\textit{Tetrabiblos} 1.5.1 H26, R39.}

Ptolemy's \textit{post hoc} rationalisation of existing astrological lore shows that certain overarching conceptual associations for planets were so widely accepted as to be indisputable. These broad meanings of each planet were likely known even to non-astrologers in Valens' time.\footnote{The \textit{Satyrica} of Petronius (late 1st century CE) includes a scene where the freedman Trimalchio, showing off at a dinner party, lists and identifies various astrological characteristics of people born under different planets and signs of the zodiac. Some of these have good overlap with predictions found in astrological books – e.g. Trimalchio's identification of those born under Sagittarius as `cross-eyed' (\textit{strabones}), \textit{Satyrica} 40 \autocite{PetroniusArbiter2009PS1:}. This compares neatly to Firmicus Maternus who also claims visual problems for those born under Sagittarius, cf. \textit{Mathesis} 8.27.1; 11 \autocite{iulius_firmicus_maternus_libros_1968}.} Yet despite this they were of limited use for astrology, since in practice any horoscope contains a specific configuration of planets. Due to the length of the periods of planets, many configurations never occur in the lifetime of any given astrologer. Deciphering the configuration's precise meaning is a complex art. This is part of the paradox of astrology: whereas the meaning of each component part may be intuitive, and (assuming a horoscope can be accurately cast) the data all on display, the overall meaning can remain hopelessly obscure. Notably, while Ptolemy discusses the characteristics of the planets, he avoids any discussion of real horoscopes or even of planetary combinations. Indeed, he excuses himself from any responsibility to discuss planets in combination at all, writing that `[i]t is of course a hopeless and impossible task to mention the proper outcome of every combination'.\footnote{\textit{Tetrabiblos} 2.9.20 H143, R189.}  

In contrast, Valens is far more concerned with the real situations an astrologer will encounter. As stated above, he provides real horoscopes throughout the \textit{Anthologies}. And at the beginning of Book 1 he gives a list of the characteristics of each planet (1.1ff). After this, almost as a footnote, he explains that although there are `benefic' and `malefic' stars, their effects can be enhanced or mitigated depending on how they are disposed. Valens does not specify which are benefic and malefic, probably because he assumes these designations are known to his reader. He then proceeds, in contrast to Ptolemy, to enumerate the characteristics of combinations of two and three planets. It is worth considering the reason for this. Conceptually, it is common to separate the practice of astrology into the computing of horoscopes with astronomical methods and then their interpretation.\footnote{This distinction is famously made by Ptolemy at \textit{Tetrabiblos} 1.1 H3, where he distinguishes between what we might call `astronomy' and what we might call `astrology'. However, it is not clear whether other astrologers accepted this as a meaningful distinction, and the words \textgreek{ἀστρονομία} and \textgreek{ἀστρολογία} (Latin: \textit{astronomia} and \textit{astrologia}) are used interchangeably until the sixth century \autocite{HübnerWolfgang1989DBAu}.} But the interpretation of a new astrological chart is an integrated practice, drawing on a combination of knowledge: of course the practical computation of horoscopes and calculation of angles between planetary positions, but also on established interpretations of commonly observed patterns. That is, the qualitative lore of astrology is not only conceptual but inherently practical. The raw materials of astrology are not only the computed positions of planets and their individual `meanings' but also a knowledge of the effects of these planets in combination, as well as their shifting valency in specific zodiac signs or `places' of the chart. However, the combinatorial possibilities are huge, making iterating them in a pre-computer age a hopeless task. So the novice astrologer cannot be instructed in the pre-defined meaning of every chart. Viewed this way, it is understandable that astrologers naturally built up methods for progressive `chunking' of horoscopes into important and repeated features that themselves had a body of `pre-computed' associations. Valens does not just aim to provide his readers with the foundational principles of astrology. By providing a compendium of the characteristics of combinations of two and three planets, he provides a practical basis for interpretation once any given horoscope is broken down into intermediate component parts. Modern computers make combinatorial problems tractable. As historians, we can therefore bring new techniques to bear on this combinatorial `data' to better understand the logic and dynamics of astrological interpretation. 

\section{Planetary combination}

The seven planets and their `combinations' (\textgreek{συγκράσεων}) are among the first things discussed at length in Book 1 of the \textit{Anthologies}. It seems reasonable to take them as of primary importance for Valens. In this section, we first consider the significance of this term, before conducting a practical calculation of the actual possible occurrences of combinations to ground the analysis that follows.

\subsection{The concept of \textit{synkrasis}}

It is important at this point to demarcate exactly what Valens means by \textit{synkrasis} (\textgreek{σύγκρασις}). Philological evidence shows that the term was used beyond astrology, with the LSJ giving `a mixing together, commixture, blending, tempering' as the first sense (p.1666). The majority of uses of \textgreek{σύγκρασις} are theological or ontological.\footnote{cf. e.g. Irenaeus \textit{Against Heresies} 1.1 \autocite{irenaeus_heresies}, Hippolytus \textit{Refutation of All Heresies} 6.30.4, both in reference to the theology of the Valentinian so-called `gnostics' \autocite{hippolytus_rofallh}; cf. also Plato, \textit{Philebus} 64d on mixed being \autocite{platophilebus}.} Among astrological writers \textgreek{σύγκρασις -εως} appears in Valens,\footnote{e.g. \textgreek{σύγκρασις} at \textit{Anthologies}  1.20; 2.28; 4.4; 4.11. \textgreek{σύγκρασιν} at 7.4. }  Ptolemy,\footnote{e.g. \textgreek{σύγκρασιν} at \textit{Tetrabiblos} 1.3.3; 1.3.15; 2.9.19; 2.9.20.} Hephaestio of Thebes,\footnote{e.g. \textgreek{σύγκρασις} at \textit{Apotelesmatica} p.333 l.18. \textgreek{σύγκρασιν} at p.51 l.4, l.7; p.172 l.7; p.332 l.33. \autocite{pingree_hephaestionII}. See also \textcite[97]{pingree-hephaestionI}.}  ps-Manetho,\footnote{e.g. \textgreek{σύγκρασις} at \textit{Apotelesmatica} 2.400; \textgreek{σύγκρασιν} at 3.227. For more on ps-Manetho, see \textcite{ps-manetho_apotelesmatica_2020}.}  Paulus Alexandrinus,\footnote{e.g. \textgreek{σύγκρασις} at \textit{Elementa Apotelesmatica} p.68 l.2. \autocite{paulus_alexandrinus}}  in a commentary by Olympiodorus on Paulus Alexandrinus,\footnote{e.g. \textgreek{σύγκρασις} at \textit{Commentarium in Paulum Alexandrinum} p.73 l.6. This work, formerly attributed to Heliodorus, is now widely credited to Olympiodorus. See \textcite[428-9]{Sphujidhvaja1978TYoS}. See also \textcite{heliodorus_boer}.}  and occasionally in critiques of astrology, e.g. Origen of Alexandria.\footnote{e.g. \textit{Philocalia} 23.18. \autocite{robinsonphilocalia}.} All of these uses refer to the mixing of the effects of planets, with some uses making more abstract reference to the planets in general (ps-Manetho, Hephaestio), and some uses referring more specifically to the mixing of particular planets. A typical use is Paulus of Alexandria referring specifically to the mixing of the effects of Mars and Mercury in the place of Good Fortune, i.e. the fifth house.\footnote{See above, \textit{Elementa Apotelesmatica} p.68 l.2.}

Scholars have argued that the concept of \textit{synkrasis} was an important one in astrology. For example, in his notes on the Corpus Hermeticum, Festugière notes that the astrological doctrine of \textit{synkrasis} is well known (\textit{bien connue}) and the term is used to designate both the combination of elements (or elementary qualities) in the body and the combination of influence of the stars.\footnote{\textcite[88]{festugiere_corpus_1954}} Ptolemy uses the word in both senses in the \textit{Tetrabiblos} (the body: 1.2.18 H11; the stars: e.g. 2.9.18 H142) suggesting a strong analogy between his understanding of how planetary influences combine in an Aristotelian framework. Valens' use of the word in the \textit{Anthologies} applies only to planets. As we show below, quantitative analysis of the characteristics he assigns to planetary combinations suggests Valens is also using a (less explicit but nevertheless present) underlying conceptual framework. 

The use of \textit{synkrasis} suggests that Valens is not talking specifically about conjunction (the occurrence of two or more planets in the same sign of the zodiac) but more generally about how the astrologer should take the influences of two or more planets into account together in a horoscope. These influences must be modulated by the particular aspect of the planets. How exactly the influences of planets combined was a practical question. In astrology, one of the most important possible combinations was conjunction: the occurrence of two planets in the same sign of the zodiac. Planetary combinations were discussed using the term `aspect': the geometrical relationship between the positions of planets in the zodiac which translated into a conceptual schema of combined influence. While conjunction was often thought of as its own category, it is sometimes counted alongside the other aspects too.\footnote{Valens sometimes distinguishes lexically between conjunctions and aspects, e.g. \textit{Anthologies} 1.1.44 \textgreek{τῆς ἑκάστου ἀστέρος μαρτυρίας ἢ συμπαρουσίας} However, conceptually, conjunction often plays a similar role to the other aspects.} Other important aspects include the benefic aspects of trine and sextile, and the malefic aspects of opposition and square.\footnote{The aspects are geometrical relationships between planets and signs, conceived of as the planets literally looking at one another. \textit{Conjunction} is the most basic aspect (same sign). \textit{Opposition} is defined as planets at opposite points on the zodiac, e.g. one in Aries and the other in Libra (six signs apart). \textit{Square} is defined as planets a quarter of the way round the zodiac from each other e.g. one in Aries and the other in Cancer (three signs apart). \textit{Trine} is a third of the way round the zodiac e.g. one planet in Aries and another in Leo (four signs apart). \textit{Sextile} is a sixth of the zodiac e.g. one planet in Aries and one in Gemini (two signs apart).} When Valens discusses \textit{synkrasis}, we believe he has all of these aspects in mind, and is providing a framework for the astrologer to modulate the predicted outcomes appropriately according to whether the aspect in question is benefic or malefic and the relative strength of each planet's influence.

\subsection{Practical limits on planetary combination}

Whether or not astrology is viewed as a science, it rests upon a foundation of empirical data about the sky. It is therefore useful to consider the constraints on planetary combination that arise in real astrological data. First, the calculation of combinations of the seven planets and luminaries; in his work, Valens considers both combinations of two planets (`doubles') and combinations of three planets (`triples'). The number of combinations of n non-repeated elements from seven elements is given by ${7 \choose n}$ (`seven choose n'). There are thus ${7 \choose 2}$=21 possible doubles and ${7 \choose 3}$=35 possible triples. In the \textit{Anthologies}, the descriptions of planetary \textit{synkrasis} omits two doubles (Mars Sun; Mars Moon), and three triples (Saturn, Mercury, Sun; Saturn, Mercury, Moon; Saturn, Sun, Moon). Although omission in the original cannot be ruled out, it would be surprising for a handbook aiming to be a comprehensive and practical guide, and we therefore suggest it may have arisen from a simple transcription error.\footnote{The fact that the combinations are similar suggests they may have been in the same section of the manuscript at one point; however, this could also reflect their being overlooked by Valens originally. The \textit{Anthologies} has a complex manuscript history which makes this kind of error very plausible, cf. \textcite{pingree_byzantine_1983}.} 

Importantly, not all combinations of planets are realised, particularly once one brings in the signs of the zodiac as well. A scheme outlined by \textcite{boxer_scheme_2020} can be used as a simplifying way to discuss the historical occurrences of conjunctions. In brief, the twelve signs of the zodiac can be represented in hexadecimal notation.\footnote{The twelve signs of the zodiac are represented by the digits 0-9 and the letters A and B as follows: Aries (0), Taurus (1), Gemini (2), Cancer (3), Leo (4), Virgo (5), Libra (6), Scorpio (7), Sagittarius (8), Capricorn (9), Aquarius (A) and Pisces (B).} The current zodiacal sign of each the seven planets (in the order Saturn, Jupiter, Mars, Sun, Venus, Mercury, Moon)\footnote{As noted previously, this is not in fact the order Valens typically uses; he usually places the luminaries at the start.} can then be converted into a digit in hexadecimal notation, so a configuration of planets can be represented as a seven-digit code. We follow Boxer in referring to this as a Z-code (from `Zodiac-code').\footnote{See \textcite[141-148]{boxer_scheme_2020}. The average persistence of a Z-code in time as calculated by Boxer is ~1.75 days, with each lasting longer than a day and no longer than 2.6 days.} For example, the Z-code

\begin{quote}
    \centering AA1A9A0
\end{quote}

means that Saturn is in Aquarius (A), Jupiter in Aquarius (A), Mars in Taurus (1), the Sun in Aquarius (A), Venus in Capricorn (9), Mercury in Aquarius (A), and the Moon in Aries (0). Z-codes offer a convenient way to simplify historical astronomical data to investigate the planetary combinations realised within the resulting horoscopes.\footnote{\label{wholesignsfootnote} It is a simplification. Rather than using presence in the whole sign to define conjunction, astrologers would use definitions based on `orbs' around a planet's exact longitude of different degrees {--} \textcite[37]{PaulusAlexandrinus+2012} gives four different orbs in his summary of earlier doctrines ($3^\circ, 7^\circ, 15^\circ, 30^\circ$). Two planets having the same sign in a Z-code is a more permissive definition of `conjunction' than the narrower orbs. The same logic extends to other aspects. (We are grateful to a reviewer for this point.) However, qualitative patterns will still hold for an arbitrarily more restrictive definition i.e. if two planets A and B are more often in conjunction than C and D by the `same sign' definition,  A and B will also be more often in conjunction than C and D by a more restricted definition.} This is appropriate for Valens, since he usually quotes only the zodiacal signs for each planet.\footnote{\textcite[89]{neugebauer_greek_1959} suggest that Valens usually computed horoscopes with an accuracy of degrees and minutes but reported only the signs. Neugebauer and Van Hoesen list the horoscopes where Valens explicitly refers to degrees and minutes as: L 110,III (p. 105), L 114,V (p. 110), L 127,XI (p. 123). The horoscopes explicitly referring to degrees only are: L 61,X (p. 82-83), L 74,IV (p. 85-86), L 75,1 (p. 89-90) and L 115,II (p. 112). Page references are to discussion of each horoscope in \textcite{neugebauer_greek_1959}.} 

We used Flatlib, a Python 3 library for traditional astrology\footnote{See \texttt{https://flatlib.readthedocs.io}} in conjunction with the Swiss Ephemeris to calculate the Z-code for each day in Valens' century and the century before, from 1st January 1 CE to 31st December 200 CE, taking midday on each day. We used the latitude and longitude of Alexandria 31.22$\degree$ N, 29.96$\degree$ E.\footnote{See \textcite[183]{neugebauer_greek_1959}.} One should expect slightly different results with this approach from what would have been calculated by contemporary astrologers such as Valens at the time due to a different frame of reference.\footnote{See \textcite[179-182]{neugebauer_greek_1959}. We are grateful to a reviewer for highlighting this point. Here we are analysing overall patterns rather than attempting to link horoscopes to specific days.} From these Z-codes, we found the days which included conjunctions. As noted above, this approach treats all cases where planets share the same sign as conjunctions (see \ref{wholesignsfootnote}). 

One immediate observation is that planetary conjunctions are extremely common: 97.5\% of days in this period contain at least one conjunction.\footnote{The expectation for random Z-codes is slightly lower at around 90\% (see Appendix).} 74.3\% contain a double conjunction, 34.4\% contain a triple conjunction, and 17.2\% contain both a double and a triple conjunction (each in a different sign). Even quadruple conjunctions occur on 7.5\% of days, although higher order conjunctions are much rarer (n=5, 0.85\%; n=6, 0.07\%; n=7, 0.0\%). 

These results demonstrate that planetary conjunctions occur in the vast majority of possible horoscopes. Even if a concept of \textit{synkrasis} was only important for conjunction, it would be invoked at least once in the interpretation of almost every horoscope. In fact, when other aspects are taken into account, any horoscope contains at least nine aspectual relationships between planets (see Appendix).\footnote{Since there are $7 \choose 2$=21 possible pairwise relationships, this means more than $40\%$ of pairwise connections between planets in a horoscope with this system of aspects will involve having to account for the double combinations of their influences (\textit{synkrasis}).} Valens' notions of planetary combination are therefore of immense value for understanding how practising astrologers conceptualised the combinatorial mainstays of horoscopic prediction.  

\subsection{The length of Valens' descriptions}

\begin{figure*}
\centering
  \includegraphics[width=0.75\textwidth]{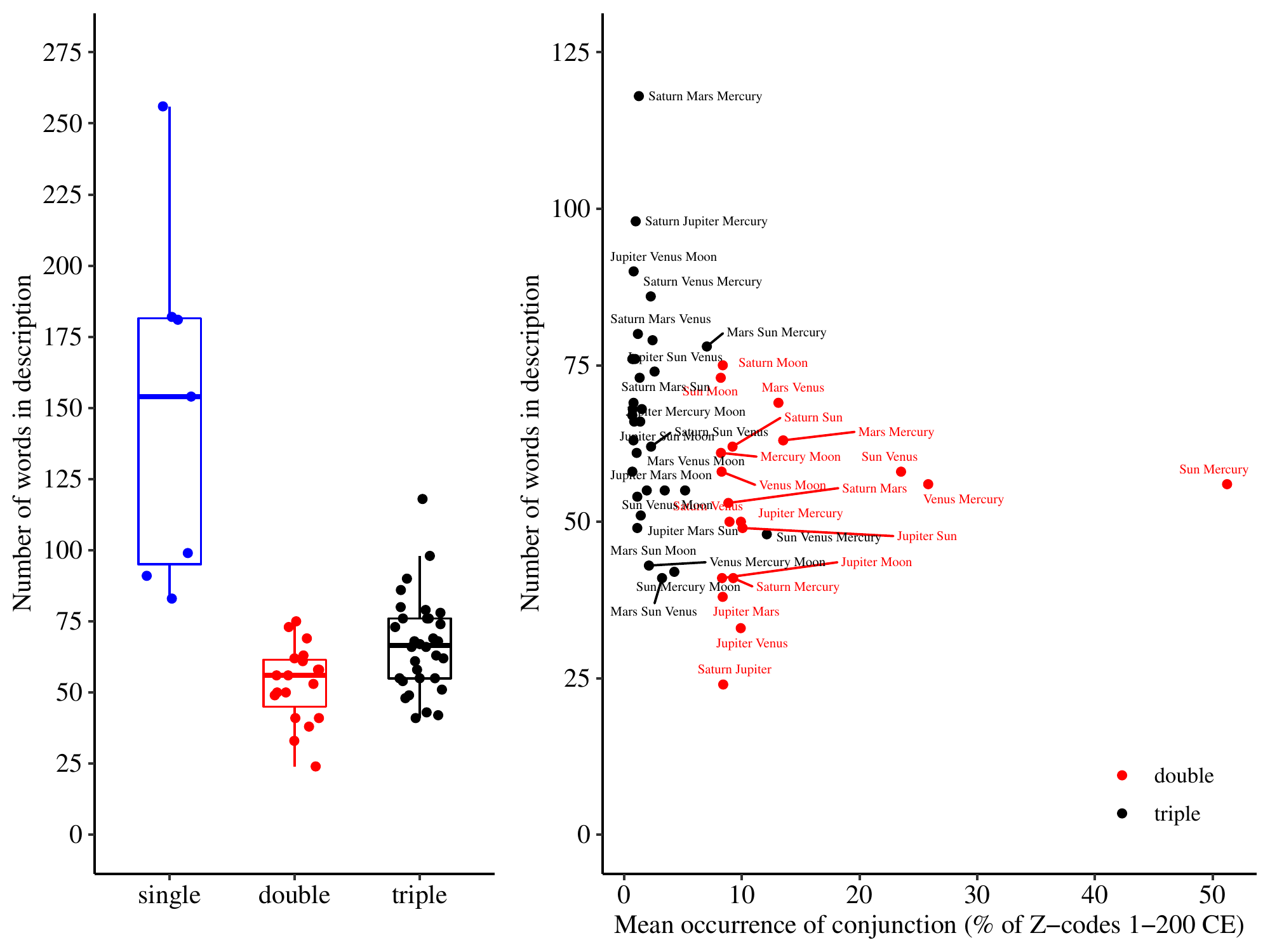}
\caption{The length of descriptions suggests conceptual importance. Left: Valens writes longer descriptions of singles than of combinations on average, but devotes more length to triples than to doubles. Right: there is no relationship between the occurrence of a conjunction of the bodies involved in a conjunction and the number of words Valens uses to describe its qualities. N.B. The y-axes are of a different scale between plots.}
\label{fig1}      
\end{figure*}

The length of the descriptions alone already gives some indication of Valens' priorities (Figure \ref{fig1}). Notably, Mercury has the longest description of any planet; as the fastest-moving, it frequently interacts with other planets, so its modifying character is important for Valens. However, particular combinations involving Mercury are not particularly notable since they recur frequently; for example, the Sun and Mercury will be in conjunction in over 50\% of horoscopes. While the cumulative number of words devoted to triples is the largest (planets 1,046; doubles 1,010; triples 2,145), Valens devotes more space on average to single planets than to combinations (mean number of words in descriptions: planets 149; doubles 53; triples 67) However, some combinations have longer descriptions than some planets. For example, the triple `Saturn Mars Mercury' has the longest description of any combination, consistent with its conceptual importance: for Valens, the combination of two malefic planets with the modifier of Mercury requires a great deal of exposition. 

There is no overall relationship between the frequency of occurrence of a conjunction of planets with the length of the description of the corresponding combination (Figure \ref{fig1}). Although triple conjunctions occur less frequently than double conjunctions, Valens devotes more space on average to describing triple combinations. This may suggest that he attaches more importance to elucidating the characteristics of triples, which involve the mingling of three influences and so are correspondingly more complex combinations than doubles.

\section{Sentiment analysis}

As noted above, the planets in Greek astrology had associated characteristics. Perhaps the most fundamental was their categorisation as benefic (positive) or malefic (negative). We decided to therefore attempt to quantify the descriptive terms used by Valens as positive, negative, or neutral, to provide a rough quantitative index as to the overall `sentiment' of a particular planet or planetary combination. 

Sentiment analysis is a method for analysing snippets of text for positive or negative valences.\footnote{For an introduction, see \textcite{CardieClaireOMaS}.} Some types of sentiment analysis use natural language processing software, while others rely on manually-mined data. Some approaches use both, especially with complex texts such as literary works or historical sources.\footnote{For a discussion of the limits of sentiment analysis, see \textcite[86-99]{DobsonJamesE2019CDHT}. An example that uses both approaches on a text from classical antiquity is \textcite{yeruva-etal-2020-interpretation}.} Here, we used a manual approach to break sentences within descriptions into lexical terms and assess their individual sentiments, and then compute mean overall sentiments for each description. 

First, we manually sorted Valens' descriptions of the planets and the effects of their combinations from the following sections of the \textit{Anthologies}:

\begin{quote}
\textit{Anthologies} 1.1 - a list of the characteristics and effects of each planet.

\textit{Anthologies} 1.19 - a list of the effects of each double combination.

\textit{Anthologies} 1.20 - a list of the effects of each triple combination.
\end{quote}

\subsection{Computing a sentiment index}

Sentences in Valens' passages of planetary descriptions tend to appear in one of four different forms. The initial descriptions of each single planet appear primarily in the following form:

\begin{quote}
    Planet $X$ is $A$, $B$, $C$ (where $A$, $B$, $C$ are adjectives).
\end{quote}

For the descriptions of the effects of planets, either singly or in combination, terms appear primarily in the three following forms: 

\begin{quote}
    $X$ causes or signifies $D$ (where $D$ is an abstract noun e.g. `death' or `illness').\footnote{Ancient philosophers debated intensively on the question of whether the stars only \textit{signify} events on earth or whether they also \textit{cause} . See e.g. \textcite{allen_greek_2010}. The question of Valens' stance on this is beyond the scope of our article.}
\end{quote}

or 

\begin{quote}
    $X$ causes or signifies [those linked to it to be] $E$s (where $E$ is a noun that encodes a practice or profession e.g. `murderers' or `goldsmiths').
\end{quote}

or 

\begin{quote}
    $X$ causes or signifies $F$s to be $G$ (where $F$ is a noun e.g. `men' and $G$ is an adjective e.g. `famous' or `wealthy').
\end{quote}

We constructed a database of terms, including the original Greek and an English translation of all nouns and adjectives used in any of these formulations, and which planet or combination of planets they applied to. One of us (Hall) then assigned to each individual term (henceforth referred to as `lexical term') a rating of `positive', `neutral', or `negative'. In many cases the assignations were very clear, due to the extremely frequent use of words with obvious sentiments like `injury' or `happiness'. In some cases, it was necessary to use cultural knowledge of Hellenistic and post-Hellenistic Greek attitudes to money, sex, certain professions, certain family arrangements and so on in order to reach a judgement on whether a term would be viewed as positive or negative.\footnote{These judgements were formulated with reference to a number of comparisons and sources. An excellent comparison can be made between Vettius Valens and his near-contemporary, the dream interpreter Artemidorus, who also wrote a long divinatory text with many specific predictions. Study of the cultural meaning and importance of Artemidorus' references is extensive, cf. e.g. \textcite{thonemann_ancient_2020} and \textcite{price_future_1986}.}  

This manual approach has several limitations. First, it can be difficult to extract lexical terms from descriptions: should `associated with fame and fortune' be two lexical terms, or just one? Second, even classifying as positive, negative or neutral requires a subjective judgment. Third, these categories are crude and do not capture any nuance of interpretation. Fourth, we only used one assessment rather than multiple expert assessments. However, despite these limitations the resulting sentiment indices are useful for comparing descriptions and looking at relationships between them. We do not put too much weight in the significance and categorisation of individual lexical terms; classifying them in this way is a means to an end, to compute a single index for each description. 

An alternative approach would be to use automated sentiment analysis with an existing tool. We explored this approach and ran descriptions (in English) through an existing sentiment analysis tool in the Stanza Python 3 library.\footnote{See \texttt{https://stanfordnlp.github.io/stanza}.} This tool is trained on modern English text sources and assigns a sentence as negative, neutral, or positive. We used the mean sentiment of the sentences in a description. We found an approximate correlation (Pearson's $r$>0.6) between our manual assessments and these automatically computed sentiments. On inspection, the automated sentiment analysis appeared inferior to our assessment. As an indicative example, take the following sentence:

\begin{quote}
    Men also derive benefits from women, and coming into possession of estates and land, they become lords. (Saturn, Jupiter, the Moon)
\end{quote}

In the context of astrological prediction, this sentence is uncomplicatedly positive. We assessed it as having two positive lexical terms: `derive benefits from women' and `coming into possession of estates and land(, they become lords)'. However, this sentence is given a neutral overall sentiment by Stanza's classifier.\footnote{It is unreasonable to expect a classifier trained on sentences such as `Yes, it's annoying and cumbersome to separate your rubbish properly all the time' (from \texttt{https://github.com/peldszus/arg-microtexts}, one of the data sources used in Stanza's model) to perform well for the purposes of analysing an astrological text translated from Greek.} Reliable assessments depend strongly on the cultural context, whether this is by training on a carefully selected training dataset or through expert knowledge. In the absence of an appropriate automated tool for Valens' context, we therefore chose to use our manual assessment to compute overall sentiment of descriptions.\footnote{We are grateful to an anonymous reviewer for the suggestion to explore existing tools. See github for more details. We found that running the analysis of the sentiment of triple combinations in terms of mean sentiments of component parts using automated sentiments recovered a much weaker correlation than in Figure \ref{fig6} (Pearson's $r\approx0.4$) and did not find that component doubles were an improvement on component singles.}

\subsection{Sentiment index results}

For the single planets, the descriptive passages range in length from 99 to 256 words, with a median of 57 lexical terms in a description. Around half of terms in the descriptions were assigned as neutral (188/400, 47\%). There is a very slight excess of positive terms compared to negative terms (29\% vs. 24\%). The malefics (Saturn and Mars) both have strong negative overall sentiment according to this method. It is of interest that the extremity of sentiment (whether positive or negative) corresponds approximately to the ordering of the five non-luminary planets: Saturn and Jupiter are equal and opposite, Mars and Venus are similarly comparable but with reduced magnitude, and Mercury has the smallest sentiment.\footnote{Speculatively, this could be because the slower-moving of the five non-luminary planets are considered to be more powerful: their relationships to other bodies change more slowly, so they have primacy. The most extreme characteristics come from  the planets which are slower-moving, modulated by the faster planets. This analysis excludes the luminaries, which do not fit neatly into this pattern but are clearly distinct in their behaviour anyway.} 

From these assignations, we then computed an overall sentiment S for each of the descriptions from their component lexical terms as

\begin{equation}
    S = \frac{p-n}{p+n+m}
\end{equation}

where p is the number of positive lexical terms, n the number of negative terms, and m the number of neutral terms. This sentiment S can range from -1 (all lexical terms are negative) to +1 (all lexical terms are positive). Table \ref{tab1} gives these values for all single planets.

For example, Saturn has 4 positive lexical terms, 21 neutral, and 50 negative. So we have

\begin{equation}
    S_\mathrm{Sa}=\frac{4-50}{4+21+50}=-\frac{46}{75}=-0.61
\end{equation} 

In the same way, we assigned lexical terms within the descriptions of double and triple combinations and then computed an overall sentiment for each description (see Supplementary Tables: Table \ref{tab2} for doubles and and Table \ref{tab3} for triples). Interestingly, the number of neutral terms was much lower for double and triple combinations, with none in descriptions of doubles and only 29 of 635 terms for triples ($<5$\%). This suggests a different conceptual categorisation for the qualities of combinations of planets as opposed to the qualities of planets on their own.

\begin{table}
\caption{Results of the sentiment assessment of the descriptions for each planet. }
\label{tab1}       
\centering
\begin{tabular}{lllllll}
\hline\noalign{\smallskip}
Planet & No. of words (Greek) & $p$ & $m$ & $n$ & Total terms & Sentiment \\\hline
Saturn & 182 & 4 & 21 & 50 & 75 & -46/75 (-0.61)\\
Jupiter	& 91 & 24 & 16 & 0 & 40 & 24/40 (0.60)\\
Mars & 154 & 8 & 25 & 39 & 72 & -31/72 (-0.43)\\
Sun & 83 & 18 & 12 & 0 & 30 & 18/30 (0.60)\\
Venus & 181 & 27 & 30 & 0 & 57 & 27/57 (0.47)\\
Mercury & 256 & 23 & 61 & 6 & 90 & 17/90 (0.19)\\
Moon & 99 & 11 & 23 & 2 & 36 & 9/36 (0.25)\\\hline
\textbf{Total} & 1,046 & 115 & 188 & 97 & 400 & 18/400 (0.05)\\
\end{tabular}
\end{table}

\begin{figure*}
\centering
  \includegraphics[width=0.75\textwidth]{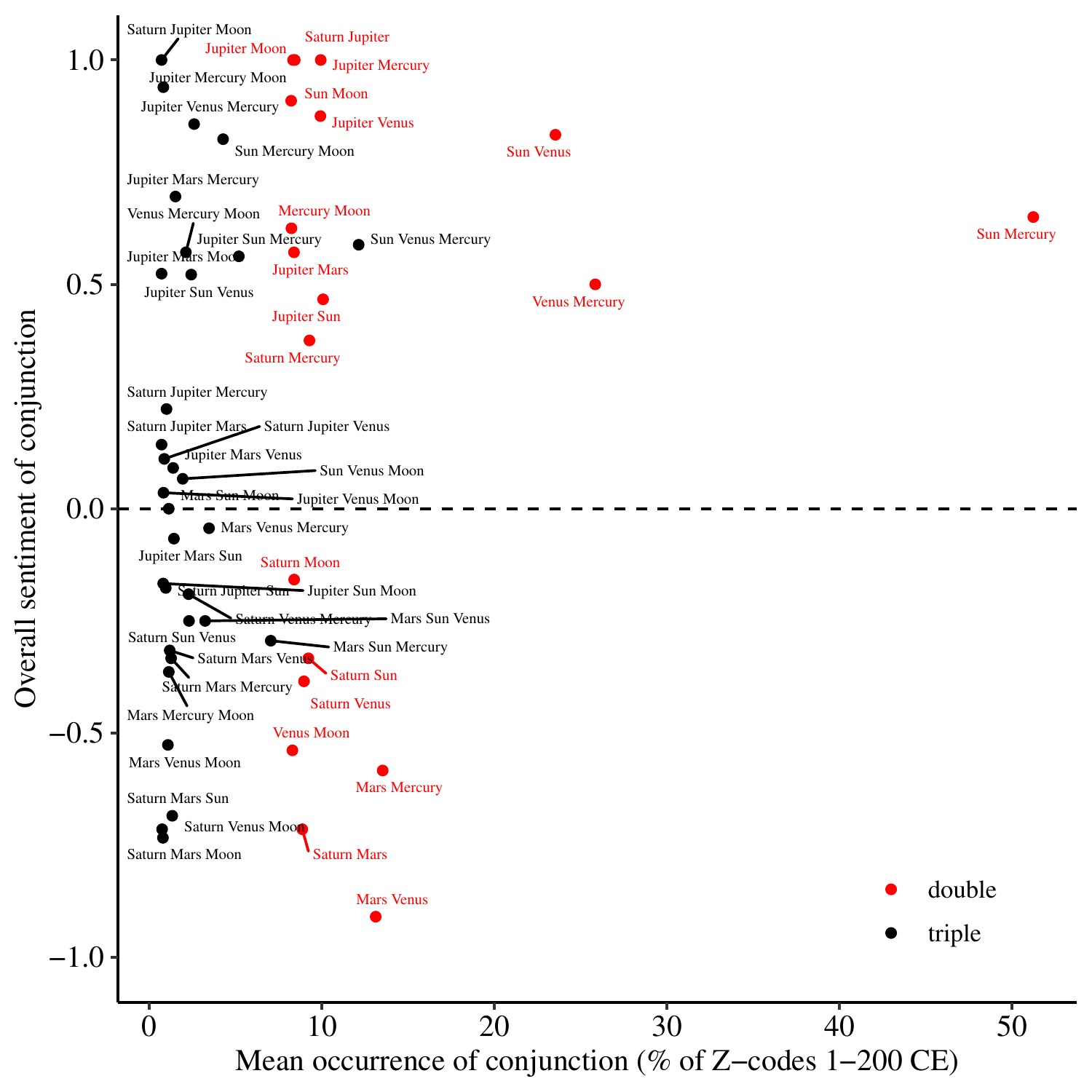}
\caption{The overall sentiment of a combination and the mean occurrence of the associated conjunction. There is no overall relationship.}
\label{fig3}      
\end{figure*}

There is no relationship between the sentiment of combinations of planets and their frequency of occurrence in conjunction (Figure \ref{fig3}). Most conjunctions occur in at most ~10\% of horoscopes, with triple conjunctions obviously rarer than double conjunctions. The three most frequent double conjunctions are those involving the Sun, Venus, and Mercury, all of which are present in more than 20\% of horoscopes. These doubles are all strongly positive in sentiment. This may suggest a certain degree of pragmatism for the working astrologer who would encounter these conjunctions in more than a fifth of their client-work. Other pragmatic considerations about the ways in which \textit{synkrasis} would be encountered might also be relevant. For example, Mercury and the Sun cannot ever be in any other aspect other than conjunction.\footnote{From a modern understanding of the solar system, this is easy to understand: the inferior planets are between Earth and the Sun, so when viewed from Earth the positions they can occupy with respect to the Sun is tightly constrained compared to the positions of the superior planets.} In a practical sense this might mean that the \textit{synkrasis} of Mercury and the Sun does not have to be as polyvalent as for most other combinations of planets. The concept of frequency of encounter of an aspect is also worth considering. Although a planet's speed is irrelevant to the number of horoscopes in which it is in aspect with another, it affects the diversity and duration of those aspects. For example, because Mercury moves quickly, in a time period of a year it will have been in every physically possible aspect with every other planet at some point during the year. In comparison, aspects between two slow-moving planets change less frequently but last for longer when they occur. So one might speculate that any given instance of \textit{synkrasis} involving Mercury would be more likely to be `routine' for an astrologer, resting on familiar knowledge, whereas a previously unencountered aspect of Saturn might prompt a consultation of existing lore to aid interpretation.

\subsection{Building combinatorial models to predict sentiment}

By reducing the descriptions of the planets and combinations down to a dataset of sentiment indices, we can perform statistical analysis to predict the sentiment of a combination. There are several options for doing this. 

\textbf{Independent bodies.} In the first instance, we used a linear model to predict S with a dummy variable xi for the presence of each body i, which takes the value 1 when body i is present in a conjunction and is 0 otherwise. That is,

\begin{equation}
    S = \sum_i \beta_i x_i
    \label{eq:dummy}
\end{equation}

\begin{figure*}
\centering
  \includegraphics[width=0.75\textwidth]{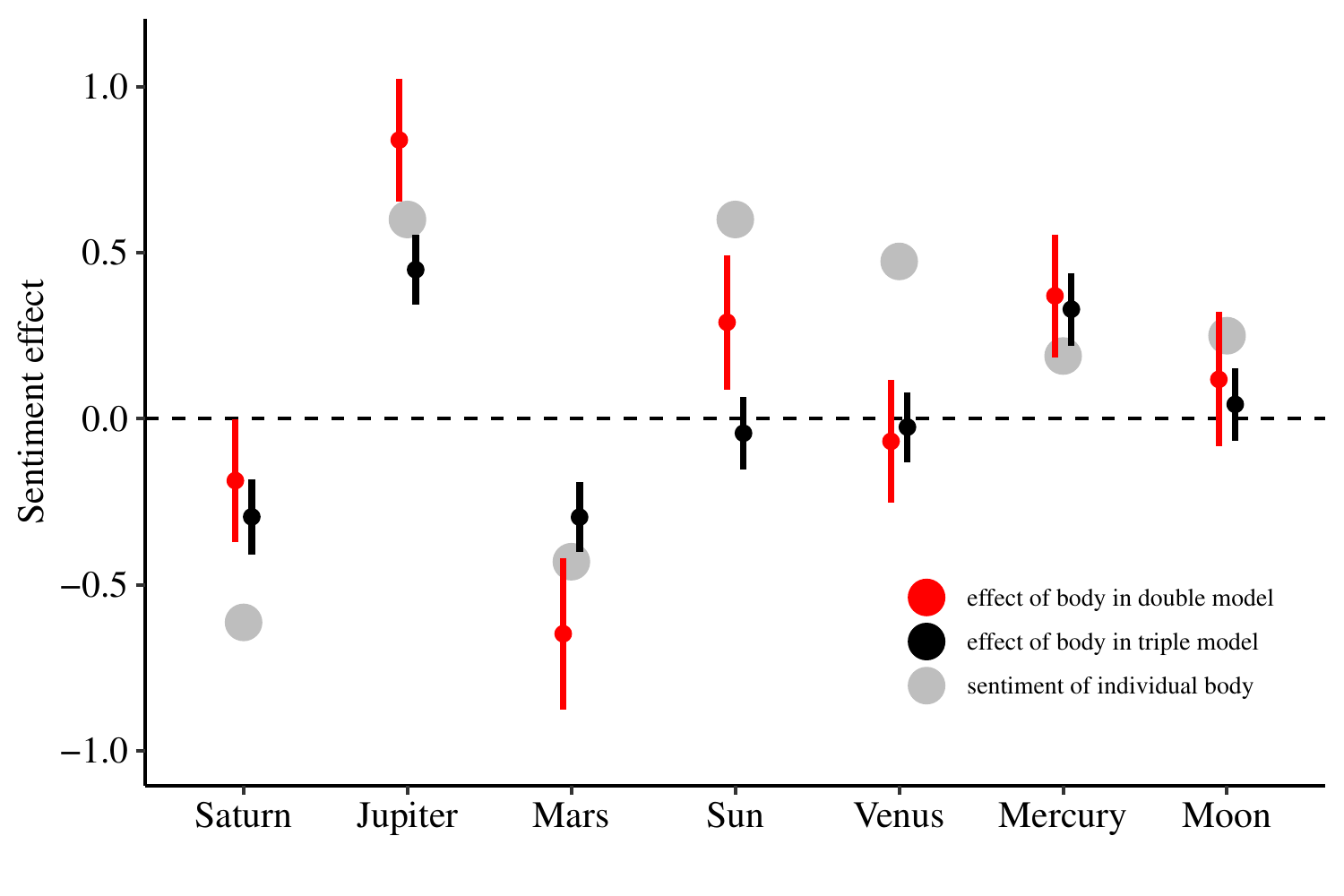}
\caption{The effects of each planet in different models. Fitted models (using eq. \ref{eq:dummy}) to predict the sentiment of doubles (red) and triples (black). Grey points show the sentiment of the body from its individual description for comparison (see Table \ref{tab1}). Lines show +/- standard error on the coefficient estimates in each model.}
\label{fig4}      
\end{figure*}

To take a specific example, $S_{\mathrm{SaJu}} = \beta_{\mathrm{Sa}}+\beta_{\mathrm{Ju}}$. We set the intercept term to zero, allowing us to calculate an effect for each dummy variable. Intuitively, this is a model where each planet i contributes a consistent and fixed effect $\beta_i$ to all combinations it occurs in. The model allows for no interaction or modulation of this effect by the presence of other planets in the combination. We fitted a model of this type for the double sentiments and the triple sentiments separately, and also compared the effect for each planet to its individual sentiment (Figure \ref{fig4}).

The results show a good concordancy between the double (red) and triple (black) models. The effect of Saturn and Mars in a combination is overall a negative one, whereas Jupiter is the most positive predictor in both models. There are some interesting discrepancies between the sentiment of the planets and their effects in the models. Venus has a positive sentiment (it is benefic) but in fact has no average effect in combinations. The Sun appears positive in doubles, but has no effect in triples. 

\textbf{Mixing models.} Rather than assuming that each planet has an overall independent effect when it is present in a combination and calculating this effect, we can work from its single sentiment. In this approach, we assume that the sentiment of a conjunction is related to the mean of the sentiments of its component parts.\footnote{It would be possible to investigate other functions, but as an average over the component parts the mean seems an obvious choice.} For a double, the only possible component parts are single planets. Thus, the predicted sentiment $\tilde{S}_{pq}$ of planets $p$ and $q$ in combination is

\begin{equation}
    \tilde{S}_{pq} = \frac{1}{2} (S_p + S_q)
    \label{eq:model_doubles}
\end{equation}

\begin{figure*}
\centering
  \includegraphics[width=0.5\textwidth]{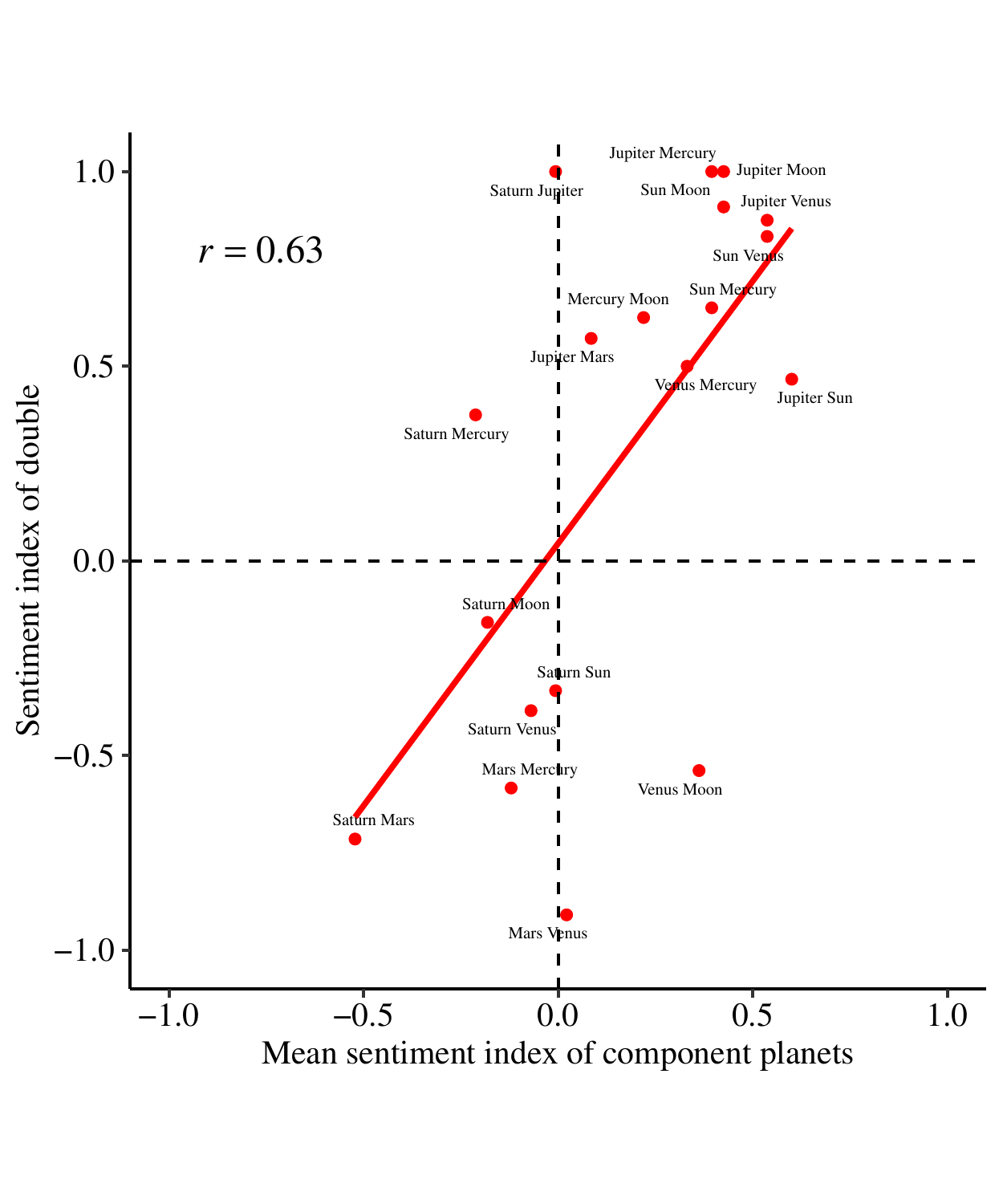}
\caption{The sentiment of double combinations is correlated with the sentiment of component planets (eq. \ref{eq:model_doubles}).}
\label{fig5}      
\end{figure*}

This predicted sentiment is strongly correlated with the actual sentiment (Figure \ref{fig5}), although there are some discrepancies. For example, the mean sentiment of Saturn and Mercury is negative, but in fact in combination they are broadly positive in effect together. Thus the relationship is not simply additive.

For the case of triples, there are multiple possible component parts. At the base level, a triple is composed of three planets, so 

\begin{equation}
    \tilde{S}_{pqr} = \frac{1}{3} (S_p + S_q + S_r)
    \label{eq:model1}
\end{equation}

However, one can also consider a triple as a combination of three doubles. In this model, one has

\begin{equation}
    \tilde{S}_{pqr} = \frac{1}{3} (S_{pq} + S_{qr} + S_{pr})
    \label{eq:model2}
\end{equation}

A third possibility is in fact given by taking into account the three possible combinations of doubles and singles:\footnote{This approach of averaging over all combinations may not always hold. For particular configurations where there is an obvious pair and an obvious singleton (for example, a conjunction and another planet in trine), this might be the simplest way to conceive of a triple; however, in many astrological settings such a designation is not possible. Given Valens provides a rather abstract list of combinations, we therefore average over all potential combinations rather than prioritising particular divisions. We choose to weight the double sentiment contribution as twice that of the single contribution, hence the factors of $\frac{2}{3}$ and $\frac{1}{3}$ respectively.}

\begin{equation}
\begin{split}
    \tilde{S}_{pqr} = \frac{1}{3} \bigg( &\big(\frac{2}{3}S_{pq} + \frac{1}{3}S_r\big) +\\
    &\big(\frac{2}{3}S_{qr} + \frac{1}{3}S_p\big) +\\
    &\big(\frac{2}{3}S_{pr} + \frac{1}{3}S_q \big) \bigg)
\end{split}
\label{eq:model3}
\end{equation}

\begin{figure*}
\centering
\makebox[\textwidth][c]{\includegraphics[width=1.3\textwidth]{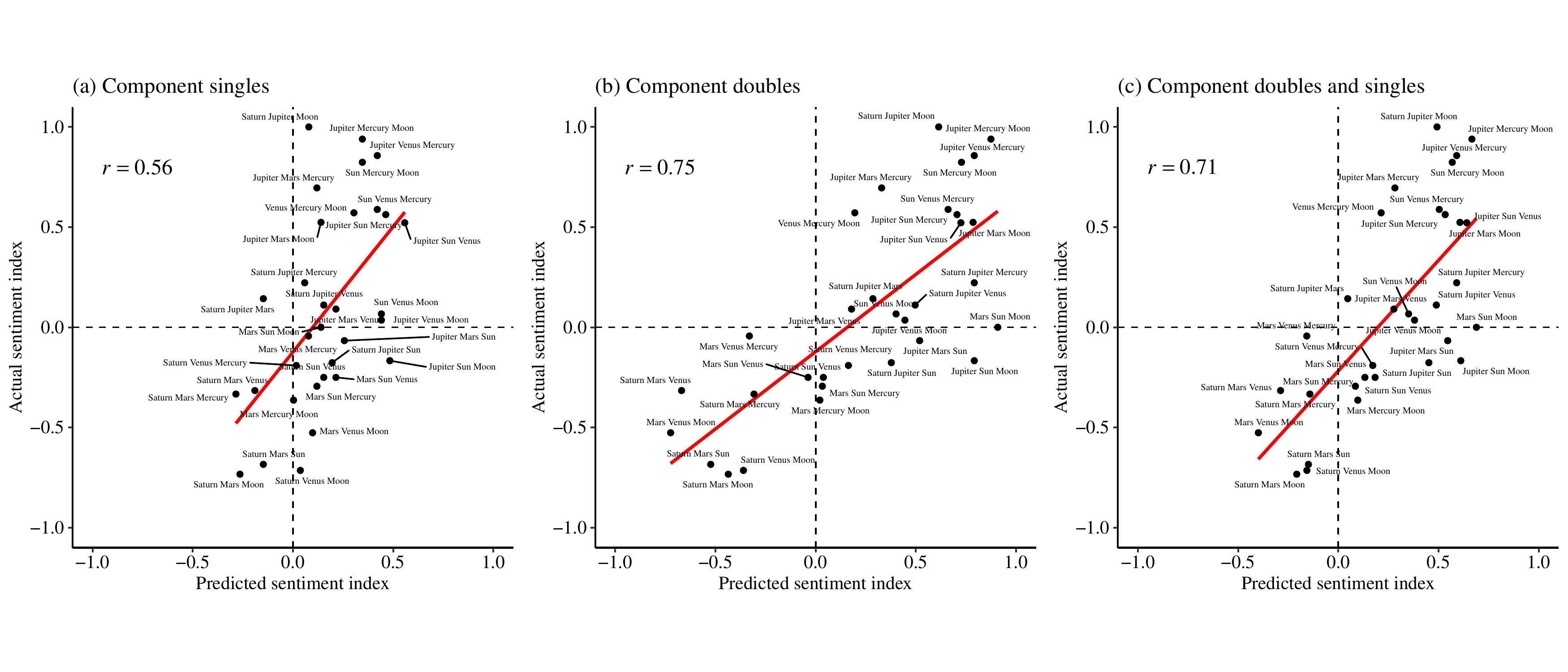}}
\caption{The sentiment of triple combinations is correlated with the mean sentiment of component parts. (a) Predicted sentiment from averaging over the three component planets (eq. \ref{eq:model1}) . (b) Predicted sentiment from averaging over the three component double combinations (eq. \ref{eq:model2}). (c) Predicted sentiment from averaging over the three component double and single combinations (eq. \ref{eq:model3}). $r$ values are Pearson's correlation coefficient.}
\label{fig6}      
\end{figure*}

Figure \ref{fig6} compares these three possible models for predicting the sentiment of the triple combinations. Pearson's correlation coefficient, which measures the degree of linear correlation between the predicted and actual sentiment, shows that a model based on single planets only is the least accurate (Figure \ref{fig6}a). A model using the mean sentiment of the doubles (Figure \ref{fig6}b) is substantially better. Using the combination of doubles and singles produces a very similar result (Figure \ref{fig6}c). We can conclude from this that the sentiment of triples is best explained by a model based on their component doubles.\footnote{Comparing these models to the `independent bodies' model using a model comparison metric such as Akaike's information criterion (AIC) also suggests that the component doubles model is the best model. AIC offers a way to compare models with different numbers of parameters for the same dataset; one chooses the model with the lowest AIC. Model AICs for Figure \ref{fig6}are: (a) 37.8, (b) 24.6, (c) 32.1, compared to the AIC of the `independent bodies' model of 28.7. Essentially, this means the `best' models for triple sentiments is the `component doubles' model. (Using other model comparisons e.g. Bayesian information criterion gives the same conclusion - see github.) Adding in more parameters for each planet as in the `independent bodies' model for the triples does not result in sufficient improvement in the fit to justify these additional parameters{---}an average based on the previously calculated (i.e. not fitted) other sentiments for doubles and singles is superior.} The internal logic of planetary combination is not simply additive but is instead combinatorial.

This may seem counterintuitive at first glance. If one was thinking of the conjunction of three planets, one might more readily imagine the combination of the three individual sets of characteristics rather than some combination of all three possible doubles. However, in other scenarios that involve the combination of three planets – for example, scenarios in which different aspects are in play – the combination of individual planets becomes much more difficult to intuit.

Take for example, the following partial horoscope: Mars in Cancer, Sun and Venus in Aries.\footnote{This horoscope does not come from Valens; it is the horoscope of one of the authors and is used for purposes of illustration.} In such an example, it is not clear how one would work in the relationships of conjunction (between Sun and Venus) and square (between Mars and Sun; between Mars and Venus) into a single calculation which combines the three individual planets. Instead, one might take either of the following routes: 

\begin{enumerate}
    \item to consider Sun and Venus as a double; to then consider Mars as an individual in square to the double Sun-Venus.
    \item to consider Sun and Venus as a conjunction-double; to consider Mars and Sun as a square-double; to consider Mars and Venus as a square-double. Then to combine the constituent doubles into a triple.
\end{enumerate}

\begin{figure*}
\centering
  \includegraphics[width=0.5\textwidth]{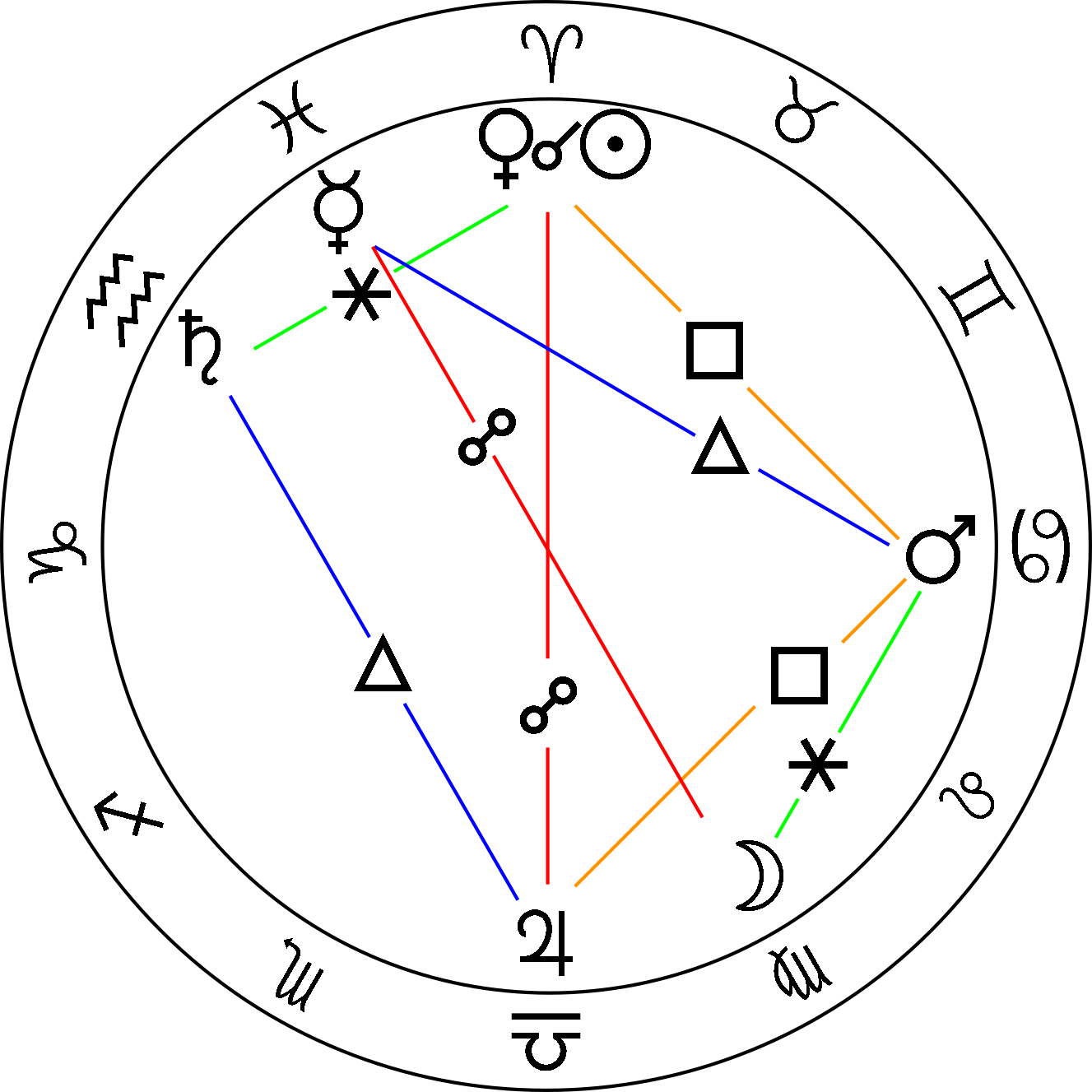}
\caption{An example horoscope with Z-code A6300B5. The outer ring shows the zodiacal signs; the inner ring shows the positions of the planets. Aspectual relationships are shown by coloured lines: opposition (red), square (orange), trine (blue), sextile (green).}
\label{fig7}      
\end{figure*}

While the first of these may seem more straightforward, the evidence from our analysis of Valens' descriptions of the combinations points more strongly towards the second. Again, the reasons for this are much clearer when considering the broader horoscopic context: such a partial horoscope would never crop up in practice since there would always be other additional planetary relationships to take into account. A typical horoscope might include aspectual combinations of four, five, or even six or seven planets; even if there appear to be some different `nodes' in the aspect network, it is extremely common for all of the planets to fit into a single linked network through a set of aspectual relationships.  In the case of our example horoscope, the rest is as follows: A6300B5, i.e. Saturn in Aquarius, Jupiter in Libra, Mars in Cancer, Sun in Aries, Venus in Aries, Mercury in Pisces, Moon in Virgo. There are twelve aspects in total:

\begin{quote}
    Saturn in trine with Jupiter; Saturn in sextile with Sun; Saturn in sextile with Venus; Jupiter square to Mars; Jupiter in opposition to Sun; Jupiter in opposition to Venus; Mars square to Venus; Mars square to Sun; Mars trine to Mercury; Mars in sextile with Moon; Sun in conjunction with Venus; Mercury in opposition to Moon.
\end{quote}

These double relationships give various options for triple combinations of influence for planets linked by aspectual relationships: one might like to think of Sun-Venus-Mars as a triple, or one might also consider Saturn-Sun-Venus or Jupiter-Sun-Venus. Quadruples made up either of doubles or triples are clearly also possible, and so on up to a septuple made up of a combination of smaller units. It is rare for astrologers to couch their predictions in the language of septuples, and one can see why it is not a natural or helpful unit to work with. 

This example serves to clarify two points. First, the combinatorial possibilities of horoscopy give the astrologer a great deal of interpretative scope: although the positions and the aspects of the stars are fixed by the input data, astrologers can choose the levels and the grouping with which to carry out analysis. Second, doubles are the natural building block in this sort of qualitative calculation based on aspects. This example helps to explain why the sentiment of Valens' triples are predicated on their constituent doubles. It also suggests a possible reason why Valens discussions of triples take up more space than his discussions of the singles or doubles{---}they may act as an exercise in showcasing the richness of this combinatorial method.

\section{Conclusion}

There is an irony in using quantitative methods in the conceptual history of astrology: the datasets are small, one is primed to look for patterns, and almost any evidence can be fitted into an existing interpretative framework. Within these constraints, we have attempted to present the evidence as we found it. Our central conclusion from investigating Valens' descriptions of \textit{synkrasis} in detail is that this planetary combination has a largely consistent internal logic, even if this is not articulated in the \textit{Anthologies}. 

We started with a discussion of the contrast between Ptolemy and Valens. Ptolemy's astrology has an underlying and well-understood logic of Aristotelian qualities and the balancing of opposites. In comparison, it is notable that Valens does not specify which planets are benefic or malefic. Yet there is clearly some variety of conceptual co-mingling in his combinations. One never encounters a planet's pure influence alone in a horoscope; this influence is always modulated by the presence of other planets. So while elucidating the effects of the planets themselves is important, it is of little practical use. Indeed, this fits the observation that the double combinations receive shorter descriptions compared to the triples. For a practical astrologer, the richness of interpretation grows with the number of planets in combination. There is no definitive answer to what the combination of three planets represents, but a list of associations can provide an overall implicit sense. We have attempted to quantify that sense here by classifying terms in descriptions as positive or negative, and using these to compute the overall `sentiment' of a description. 

However, by providing lists of combinations in the \textit{Anthologies} Valens is implicitly accepting that the astrologer need not start from first principles every time they interpret a horoscope. Instead, they can start by noting the key combinations of influences, `chunking' the horoscope into its salient features. The third sense (c.II) of \textit{synkrasis} noted by the LSJ is a `mixture, compound...but compounded so to say of both'. We have found that the combinations of three planets are better explained by averaging over their component pairs, than by averaging over the individual planets involved. This observation suggests that Valens does not think of these thrice-wise combinations of influence merely as combinations of the fundamental planets, but as combinations of combinations. This is an iterative process that one could speculatively extend: for combinations of four planets, one does not have to turn back to the descriptions of the individual planets' characteristics. Instead, one can look at the combinations of three planets it contains, then construct a new compound of influences from these descriptions. We believe our analysis reveals a conceptual and systematic subtlety not previously recognised in Valens' work. We suspect it may be a fruitful avenue along which to investigate texts from other astrological writers who are also considered to be unsystematic or miscellaneous, including Hephaestio of Thebes, Dorotheus of Sidon, and Firmicus Maternus. The main challenge we see is the time and expertise needed to manually assess the texts to enable a similar quantification. While automated analysis could circumvent this problem, this would probably require the training of specific models for this astrological context. 

There is a great deal more to Hellenistic astrology than simply calculating the combination of planets{---}for example, the complexities of the house system play an important role in Valens' work and we have hardly mentioned them here. We nevertheless suggest that this case study shows it is possible, through using statistical methods, to adumbrate conceptual systems at play in the practice of Hellenistic astrology. These conceptual systems fall neither into the modern classifications of history of astronomy (examining how ancient people calculated the positions and relations of stars and planets) nor history of philosophy (examining ancient physical theories about the stars or theories of astral causation). Instead, our method aims to capture something important yet elusive: a glimpse of the general understanding that Hellenistic astrological practitioners had of the structural coherence of their craft.


%
%


\newpage

\printbibliography

@article{riley_theoretical_1987,
	title = {Theoretical and Practical Astrology: Ptolemy and His Colleagues},
	volume = {117},
	pages = {235--256},
	journaltitle = {Transactions of the American Philological Association (1974-)},
	author = {Riley, Mark},
	date = {1987}
}

@book{PaulusAlexandrinus+2012,
author = {{Paulus Alexandrinus}},
editor = {Emilie Boer},
doi = {doi:10.1515/9783110280777},
url = {https://doi.org/10.1515/9783110280777},
title = {Eisagogika / Elementa apotelesmatica},
year = {2012},
publisher = {B. G. Teubner},
ISBN = {9783110280777}
}

@book{HeilenStephan2015,
series = {Texte und Kommentare ; 43},
booktitle = {"Hadriani genitura" - Die astrologischen Fragmente des Antigonos von Nikaia : Edition, Übersetzung und Kommentar},
isbn = {9783110288735},
year = {2015},
title = {"Hadriani genitura" - Die astrologischen Fragmente des Antigonos von Nikaia : Edition, Übersetzung und Kommentar [electronic resource]},
copyright = {Available until 6 September 2022 via De Gruyter.},
language = {ger},
address = {Berlin ; Boston},
author = {Heilen, Stephan},
}

@book{Sphujidhvaja1978TYoS,
series = {Harvard oriental series ; v. 48},
publisher = {Harvard University Press},
isbn = {9780674963733},
year = {1978},
title = {The Yavanajātaka of Sphujidhvaja},
language = {eng;san},
address = {Cambridge, Mass.},
author = {Sphujidhvaja and Pingree, David},
lccn = {76048152},
}

@book{JonesAlexander2010,
series = {Archimedes (Dordrecht, Netherlands) ; v. 23},
publisher = {Springer},
isbn = {9789048127870},
year = {2010},
title = {Ptolemy in perspective : use and criticism of his work from antiquity to the nineteenth century},
language = {eng},
address = {Dordrecht ; London},
author = {Jones, Alexander},
lccn = {2009940670},
}

@article{riley_science_1988,
	title = {Science and Tradition in the "Tetrabiblos"},
	volume = {132},
	issn = {0003-049X},
	url = {https://www.jstor.org/stable/3143825},
	pages = {67--84},
	number = {1},
	journaltitle = {Proceedings of the American Philosophical Society},
	author = {Riley, Mark},
	urldate = {2021-02-10},
	date = {1988},
	note = {Number: 1
Publisher: American Philosophical Society}
}

@article{neugebauer_study_1951,
	title = {The Study of Wretched Subjects},
	volume = {42},
	issn = {0021-1753},
	url = {https://www.journals.uchicago.edu/doi/10.1086/349279},
	doi = {10.1086/349279},
	pages = {111--111},
	number = {2},
	journaltitle = {Isis},
	shortjournal = {Isis},
	author = {Neugebauer, Otto},
	urldate = {2021-03-10},
	date = {1951},
	note = {Number: 2
Publisher: The University of Chicago Press}
}

@article{neugebauer_egyptian_1942,
	title = {Egyptian Planetary Texts},
	volume = {32},
	issn = {00659746},
	url = {http://www.jstor.org/stable/1005598},
	doi = {10.2307/1005598},
	pages = {209--250},
	number = {2},
	journaltitle = {Transactions of the American Philosophical Society},
	author = {Neugebauer, Otto},
	urldate = {2021-03-10},
	date = {1942},
	note = {Number: 2
Publisher: American Philosophical Society}
}

@book{boxer_scheme_2020,
	title = {A Scheme of Heaven},
	publisher = {Profile Books},
	author = {Boxer, Alexander},
	date = {2020}
}

@book{neugebauer_greek_1959,
	location = {Philadelphia, {PA}},
	title = {Greek Horoscopes},
	publisher = {American Philosophical Society},
	author = {Neugebauer, Otto and Van Hoesen, H. B.},
	date = {1959}
}

@book{thonemann_ancient_2020,
	title = {An Ancient Dream Manual: Artemidorus' The Interpretation of Dreams},
	isbn = {978-0-19-884382-5},
	shorttitle = {An Ancient Dream Manual},
	publisher = {Oxford University Press},
	author = {Thonemann, Peter},
	date = {2020},
	doi = {10.1093/oso/9780198843825.001.0001}
}

@book{vettius_valens_antiochenus_vettii_2010,
	location = {Leipzig},
	title = {Vettii Valentis Antiocheni anthologiarum libri novem},
	series = {Bibliotheca scriptorum Graecorum et Romanorum Teubneriana},
	pagetotal = {604},
	publisher = {{BGTeubner}},
	editor = {Pingree, David},
	date = {1986},
}

@book{ptolemy_apotelesmatica_hubner,
    location = {Stuttgart},
    year = {1998},
    title = {Claudius Ptolemaei opera quae exstant omnia. Volumen III, 1},
    Volume = {3.1},
    editor = {Wolfgang H\"{u}bner},
    publisher = {Teubner}
}

@book{HübnerWolfgang1982DEdT,
series = {Sudhoffs Archiv. Beihefte, Heft 22},
publisher = {F. Steiner},
isbn = {9783515033374},
year = {1982},
title = {Die Eigenschaften der Tierkreiszeichen in der Antike : ihre Darstellung und Verwendung unter Besonderer Berücksichtigung des Manilius},
language = {ger},
address = {Wiesbaden},
author = {Hübner, Wolfgang},
}

@book{augustine_city_II,
    title = {City of God Volume II: Books 4-7},
    editor = {William M. Green},
    series = {Loeb Classical Library},
    volume = {412},
    year = {1963},
    publisher = {Harvard University Press},
    address = {Cambridge, MA}
}

@book{iulius_firmicus_maternus_libros_1968,
	location = {Berlin/Boston, {GERMANY}},
	title = {Libros {IV} Posteriores Cum Praefatione et Indicibus Continens},
	isbn = {978-3-11-096518-6},
	url = {http://ebookcentral.proquest.com/lib/oxford/detail.action?docID=936577},
	publisher = {De Gruyter, Inc.},
	author = {Kroll, Wilhelm and Ziegler, K. and Skutsch, O.},
	date = {1968}
}

@book{crymbletechnology,
    author = {Adam Crymble},
    title = {Technology and the Historian: Transformations in the Digital Age},
    year = {2021},
    Address = {Urbana, IL},
    publisher = {University of Illinois Press}}

@book{DobsonJamesE2019CDHT,
publisher = {University of Illinois Press},
booktitle = {Critical Digital Humanities},
isbn = {0252042271},
year = {2019},
title = {Critical Digital Humanities: The Search for a Methodology},
copyright = {2019 the Board of Trustees of the University of Illinois},
language = {eng},
address = {Champaign},
author = {Dobson, James E},
}

@incollection{CardieClaireOMaS,
publisher = {Oxford University Press},
booktitle = {The Oxford Handbook of Computational Linguistics 2nd edition},
isbn = {0199573697},
date = {2017},
title = {Opinion Mining and Sentiment Analysis},
language = {eng},
author = {Cardie, Claire and Breck, Eric},
}

@inproceedings{yeruva-etal-2020-interpretation,
    title = "Interpretation of Sentiment Analysis in Aeschylus{'}s {G}reek Tragedy",
    author = "Yeruva, Vijaya Kumari  and
      ChandraShekar, Mayanka  and
      Lee, Yugyung  and
      Rydberg-Cox, Jeff  and
      Blanton, Virginia  and
      Oyler, Nathan A",
    booktitle = "Proceedings of the The 4th Joint SIGHUM Workshop on Computational Linguistics for Cultural Heritage, Social Sciences, Humanities and Literature",
    month = dec,
    year = "2020",
    address = "Online",
    publisher = "International Committee on Computational Linguistics",
    url = "https://aclanthology.org/2020.latechclfl-1.17",
    pages = "138--146",
}

@book{PetroniusArbiter2009PS1:,
series = {Texte und Kommentare ; Bd. 32},
publisher = {Walter de Gruyter},
isbn = {9783110220827},
year = {2009},
title = {Petronius, Satyrica, 1-15 : Text, Übersetzung, Kommentar},
address = {Berlin ; New York},
author = {Breitenstein, Natalie}
}

@book{hippolytus_rofallh,
title = {Hippolytus. Refutatio omnium haeresium},
editor = {M. Marcovich},
publisher = {De Gruyter},
address = {Berlin},
Year = {1986},
series = {Patristische Texte und Studien},
Number = {25}
}

@book{robinsonphilocalia,
	Editor = {J.A. Robinson},
	Title = {Philocalia},
	Publisher = {Cambridge University Press},
	Address = {Cambridge},
	Year = {1913}
	}

@book{heliodorus_boer,
    title = {Heliodori, ut dicitur, in Paulum Alexandrinum commentarium},
    editor = {E. Boer},
    publisher = {Teubner},
    address = {Leipzig},
    year = {1962}
}

@book{irenaeus_heresies,
    title = {Sancti Irenaei episcopi Lugdunensis libri quinque adversus haereses},
    volume = {1},
    editor = {W.W. Harvey},
    year = {1857},
    publisher = {Cambridge University Press},
    address = {Cambridge},
    }

@book{platophilebus,
    title = {Platonis opera},
    Editor = {J. Burnet},
    volume = {2},
    year = {1901},
    publisher = {Clarendon Press},
    address = {Oxford},
    }

@book{pingree_hephaestionII,
    title = {Hephaestionis Thebani apotelesmaticorum libri tres},
    editor = {David Pingree},
    Year = {1974},
    Volume = {2},
    Publisher = {Teubner},
    Address = {Leipzig}
    }

@book{pingree-hephaestionI,
    title = {Hephaestionis Thebani apotelesmaticorum libri tres},
    editor = {David Pingree},
    Year = {1973},
    Volume = {1},
    Publisher = {Teubner},
    Address = {Leipzig}}

@book{paulus_alexandrinus,
    title = {Pauli Alexandrini elementa apotelesmatica},
    editor = {E. Boer},
    year = {1957},
    publisher = {Teubner},
    Address = {Leipzig}
}

@book{robbins_tetrabiblos,
    title = {Ptolemy: Tetrabiblos},
    editor = {F.E. Robbins},
    series = {Loeb Classical Library},
    Number = {435},
    Publisher = {Harvard University Press},
    address = {Cambridge, MA},
    year = {1940}
}

@book{HübnerWolfgang1989DBAu,
series = {Abhandlungen der Geistes- und Sozialwissenschaftlichen Klasse ; Jahrg.1989/Nr.7},
publisher = {Akademie der Wissenschaften und der Literatur ; Franz Steiner},
isbn = {9783515055901},
year = {1989},
title = {Die Begriffe "Astrologie" und "Astronomie" in der Antike : Wortgeschichte und Wissenschaftssystematik : mit einer Hypothese zum Terminus "Quadrivium"},
language = {ger},
address = {Mainz : Stuttgart},
author = {Hübner, Wolfgang},
}

@book{bouche-leclercq_astrologie_2014,
	title = {L'Astrologie grecque},
	publisher = {E. Leroux},
	Address = {Paris},
	author = {Bouch\'{e}-Leclercq, Auguste},
	date = {1899},
}

@article{freeth_model_2021,
	title = {A Model of the Cosmos in the ancient Greek Antikythera Mechanism},
	volume = {11},
	issn = {2045-2322},
	url = {https://doi.org/10.1038/s41598-021-84310-w},
	doi = {10.1038/s41598-021-84310-w},
	pages = {5821},
	number = {1},
	journaltitle = {Scientific Reports},
	shortjournal = {Scientific Reports},
	author = {Freeth, Tony and Higgon, David and Dacanalis, Aris and {MacDonald}, Lindsay and Georgakopoulou, Myrto and Wojcik, Adam},
	date = {2021-03-12},
	note = {Number: 1}
}

@incollection{greenbaum_hellenistic_2020,
	location = {Leiden},
	title = {The Hellenistic Horoscope},
	series = {Brill's Companions to Classical Studies},
	pages = {443--471},
	booktitle = {Hellenistic Astronomy: The Science in Its Contexts},
	publisher = {Brill},
	author = {Greenbaum, Dorian Gieseler},
	editor = {Bowen, Alan C. and Rochberg, Francesca},
	date = {2020}
}

@book{feke_ptolemys_2019,
	location = {Princeton, {NJ}},
	title = {Ptolemy's Philosophy: Mathematics as a Way of Life},
	publisher = {Princeton University Press},
	author = {Feke, Jacqueline},
	date = {2019}
}

@incollection{allen_greek_2010,
	location = {Chicago, {IL}},
	title = {Greek Philosophy and Signs},
	series = {Oriental Institute Seminars},
	pages = {29--42},
	number = {6},
	booktitle = {Divination and the Interpretation of Signs in the Ancient World},
	publisher = {The Oriental Institute of the University of Chicago},
	author = {Allen, James},
	editor = {Annus, Amar},
	date = {2010}
}

@article{riley_ptolemys_use,
    title = {Ptolemy's Use of His Predecessors' Data},
    author = {Riley, Mark},
    journal = {Transactions of the American Philological Association},
    Year = {1995},
    Volume = {125},
    pages = {221--250}
}

@book{nutton_ancient_2004,
	location = {New York},
	title = {Ancient medicine},
	isbn = {978-0-415-08611-0},
	series = {Sciences of Antiquity},
	publisher = {Taylor \& Francis},
	author = {Nutton, Vivian},
	date = {2004}
}

@article{thorndike_roman_1913,
	title = {A Roman Astrologer as a Historical Source: Julius Firmicus Maternus},
	volume = {8},
	issn = {0009837X},
	doi = {10.1086/359824},
	pages = {415--435},
	number = {4},
	journaltitle = {Classical Philology},
	author = {Thorndike, Lynn},
	date = {1913},
	note = {Number: 4}
}

@article{van_oort_augustines_2011,
	title = {Augustine's Manichaean Dilemma in Context},
	volume = {65},
	issn = {00426032},
	url = {http://www.jstor.org/stable/41480508},
	pages = {543--567},
	number = {5},
	journaltitle = {Vigiliae Christianae},
	author = {van Oort, Johannes},
	urldate = {2021-03-18},
	date = {2011},
	note = {Number: 5
Publisher: {BRILL}}
}

@book{taub_ptolemys_1993,
	location = {Chicago},
	title = {Ptolemy's universe : the natural philosophical and ethical foundations of Ptolemy's astronomy},
	isbn = {978-0-8126-9228-0},
	publisher = {Open Court},
	author = {Taub, Liba Chaia},
	date = {1993}
}

@collection{festugiere_corpus_1954,
	location = {Paris},
	title = {Corpus Hermeticum},
	publisher = {Les Belles-Lettres},
	editor = {Festugière, André-Jean},
	date = {1954}
}

@book{ps-manetho_apotelesmatica_2020,
	location = {Oxford},
	title = {Ps-Manetho: Apotelesmatica: Books Two, Three, and Six Introduction, Translation, and Commentary},
	publisher = {Oxford University Press},
	editor = {Lightfoot, Jane},
	date = {2020}
}

@incollection{lewis_aristotles_2018,
	location = {Cambridge},
	title = {Aristotle's Physical Theory},
	pages = {196--214},
	booktitle = {The Cambridge History of Science},
	publisher = {Cambridge University Press},
	author = {Lewis, Eric},
	editor = {Jones, Alexander and Taub, Liba Chaia},
	date = {2018}
}

@article{neugebauer_chronology_1954,
	title = {The Chronology of Vettius Valens' Anthologiae},
	volume = {47},
	issn = {0017-8160},
	url = {https://www.jstor.org/stable/1508487},
	pages = {65--67},
	number = {1},
	journaltitle = {The Harvard Theological Review},
	author = {Neugebauer, Otto},
	urldate = {2021-04-01},
	date = {1954},
	note = {Publisher: Cambridge University Press}
}

@article{riley_survey_1996,
	title = {A Survey of Vettius Valens},
	url = {https://www.csus.edu/indiv/r/rileymt/PDF_folder/VettiusValens.PDF},
	author = {Riley, Mark},
	urldate = {2021-04-01},
	date = {1996}
}

@article{riley_anthologies,
    title = {Anthologies},
    url = {https://www.csus.edu/indiv/r/rileymt/vettius%20valens%20entire.pdf},
    author = {Riley, Mark},
    urldate={2021-04-02},
    year = {2011}}

@article{price_future_1986,
	title = {The Future of Dreams: From Freud to Artemidorus},
	issn = {00312746, 1477464X},
	url = {http://www.jstor.org/stable/650978},
	pages = {3--37},
	number = {113},
	journaltitle = {Past \& Present},
	author = {Price, S. R. F.},
	urldate = {2021-03-30},
	date = {1986},
	note = {Publisher: [Oxford University Press, The Past and Present Society]},
}

@mvbook{neugebauer_history_1975,
	location = {New York, Berlin, Heidelberg},
	edition = {1st ed.},
	title = {A History of Ancient Mathematical Astronomy},
	volumes = {3},
	publisher = {Springer-Verlag},
	author = {Neugebauer, Otto},
	date = {1975}
}

@article{pingree_byzantine_1983,
	title = {The Byzantine Tradition of Vettius Valens's "Anthologies"},
	volume = {7},
	issn = {0363-5570},
	url = {https://www.jstor.org/stable/41036114},
	pages = {532--541},
	journaltitle = {Harvard Ukrainian Studies},
	author = {Pingree, David},
	urldate = {2021-04-01},
	date = {1983}
}

@incollection{greenbaum_vettius_2019,
	title = {Vettius Valens},
	rights = {Copyright © 2019 John Wiley \& Sons, Ltd},
	isbn = {978-1-4443-3838-6},
	url = {https://onlinelibrary.wiley.com/doi/abs/10.1002/9781444338386.wbeah21338.pub2},
	pages = {1--1},
	booktitle = {The Encyclopedia of Ancient History},
	publisher = {John Wiley \& Sons},
	author = {Greenbaum, Dorian Gieseler},
	urldate = {2021-04-01},
	date = {2019},
	langid = {english},
	doi = {10.1002/9781444338386.wbeah21338.pub2},
	note = {\_eprint: https://onlinelibrary.wiley.com/doi/pdf/10.1002/9781444338386.wbeah21338.pub2}
}

@article{sarton_seventy-sixth_1950,
	title = {Seventy-Sixth Critical Bibliography of the History and Philosophy of Science and of the History of Civilization (To May 1950)},
	volume = {41},
	url = {https://www.jstor.org/stable/227083},
	pages = {328-424},
	number = {125},
	journaltitle = {Isis},
	author = {Sarton, George and Siegel, Frances},
	urldate = {2021-04-01},
	date = {1950}
}

@online{koch_swiss_nodate,
	title = {Swiss Ephemeris},
	url = {https://www.astro.com/swisseph/swephinfo_e.htm},
	titleaddon = {Astrodienst},
	author = {Koch, Dieter and Treindl, Alois},
	urldate = {2021-04-01}
}

@incollection{frede_ancient_1987,
	title = {The Ancient Empiricists},
	url = {www.jstor.org/stable/10.5749/j.cttttpfd.17},
	pages = {243--260},
	booktitle = {Essays in Ancient Philosophy},
	publisher = {University of Minnesota Press},
	author = {Frede, Michael},
	date = {1987}
}

@article{kroll_review_1910,
	title = {Review of Vetii Valentis Anthologiarum Libri},
	volume = {5},
	issn = {0009-837X},
	url = {https://www.jstor.org/stable/261277},
	pages = {525--527},
	number = {4},
	journaltitle = {Classical Philology},
	author = {Moore, Clifford Herschel},
	urldate = {2021-04-01},
	date = {1910},
	note = {Publisher: University of Chicago Press}
}

@article{winlock_origin_1940,
	title = {The Origin of the Ancient Egyptian Calendar},
	volume = {83},
	issn = {0003-049X},
	url = {https://www.jstor.org/stable/985113},
	pages = {447--464},
	number = {3},
	journaltitle = {Proceedings of the American Philosophical Society},
	author = {Winlock, H. E.},
	urldate = {2021-04-01},
	date = {1940},
	note = {Publisher: American Philosophical Society}
}

@book{cumont_egypt_1937,
	location = {Brussels},
	title = {L'Egypt des Astrologues},
	pagetotal = {254},
	publisher = {Fondation Égyptologique Reine Elisabeth},
	author = {Cumont, Franz},
	date = {1937}
}

@article{macmullen_social_1971,
	title = {Social History in Astrology},
	volume = {2},
	issn = {0066-1619},
	url = {https://www.jstor.org/stable/44080033},
	pages = {105--116},
	journaltitle = {Ancient Society},
	author = {{MacMullen}, Ramsay},
	urldate = {2021-04-01},
	date = {1971},
	note = {Publisher: Peeters Publishers}
}

@article{jones_making_1941,
	title = {The Making of a Lexicon},
	volume = {55},
	issn = {0009-840X},
	url = {https://www.jstor.org/stable/704990},
	pages = {1--13},
	number = {1},
	journaltitle = {The Classical Review},
	author = {Jones, H. Stuart},
	urldate = {2021-04-01},
	date = {1941},
	note = {Publisher: Cambridge University Press}
}

\newpage

\section*{Supplementary tables}

\begin{table}[ht]
\centering
\caption{Summary of sentiment assessment for Valens' descriptions of doubles.}
\begin{tabular}{llllll}
  \hline
 Combination & Words (Greek) & $p$ & $n$ & Total terms & Sentiment \\ 
  \hline
 Saturn Jupiter & 24 & 7 & 0 & 7 & 7/7 (1) \\ 
 Saturn Mars & 53 & 2 & 12 & 14 & -10/14 (-0.71) \\ 
 Saturn Sun & 62 & 4 & 8 & 12 & -4/12 (-0.33) \\ 
Saturn Venus & 50 & 4 & 9 & 13 & -5/13 (-0.38) \\ 
Saturn Mercury & 41 & 11 & 5 & 16 & 6/16 (0.38) \\ 
Saturn Moon & 75 & 8 & 11 & 19 & -3/19 (-0.16) \\ 
Jupiter Mars & 38 & 11 & 3 & 14 & 8/14 (0.57) \\ 
 Jupiter Sun & 49 & 11 & 4 & 15 & 7/15 (0.47) \\ 
 Jupiter Venus & 33 & 15 & 1 & 16 & 14/16 (0.88) \\ 
 Jupiter Mercury & 50 & 15 & 0 & 15 & 15/15 (1) \\ 
Jupiter Moon & 41 & 15 & 0 & 15 & 15/15 (1) \\ 
Mars Venus & 69 & 1 & 21 & 22 & -20/22 (-0.91) \\ 
 Mars Mercury & 63 & 5 & 19 & 24 & -14/24 (-0.58) \\ 
Sun Venus & 58 & 11 & 1 & 12 & 10/12 (0.83) \\ 
Sun Mercury & 56 & 16 & 3 & 19 & 13/19 (0.68) \\ 
 Sun Moon & 73 & 21 & 1 & 22 & 20/22 (0.91) \\ 
Venus Mercury & 56 & 12 & 4 & 16 & 8/16 (0.50) \\ 
 Venus Moon & 58 & 3 & 10 & 13 & -7/13 (-0.54) \\ 
 Mercury Moon & 61 & 13 & 3 & 16 & 10/16 (0.62) \\ \hline
 \textbf{Total} & 1,010 & 185 & 115 & 300 & 70/300 (0.23)
\end{tabular}
\label{tab2}
\end{table}

\newpage

\begin{table}[ht]
\centering
\caption{Summary of sentiment assessment for Valens' descriptions of triples.}
\begin{tabular}{lllllll}
  \hline
Combination & Words (Greek) & $p$ & $m$ & $n$ & Total terms & Sentiment \\ 
  \hline
Saturn Jupiter Mars & 76 & 8 & 0 & 6 & 14 & 2/14 (0.14) \\ 
  Saturn Jupiter Sun & 76 & 7 & 0 & 10 & 17 & -3/17 (-0.18) \\ 
  Saturn Jupiter Venus & 66 & 10 & 0 & 8 & 18 & 2/18 (0.11) \\ 
  Saturn Jupiter Mercury & 98 & 15 & 2 & 9 & 26 & 6/26 (0.23) \\ 
  Saturn Jupiter Moon & 67 & 11 & 0 & 0 & 11 & 10/11 (0.91) \\ 
  Saturn Mars Sun & 73 & 2 & 1 & 15 & 18 & -13/18 (-0.72) \\ 
  Saturn Mars Venus & 80 & 6 & 0 & 12 & 18 & -6/18 (-0.33) \\ 
  Saturn Mars Mercury & 118 & 8 & 0 & 16 & 24 & -8/24 (-0.33) \\ 
  Saturn Mars Moon & 76 & 2 & 0 & 13 & 15 & -11/15 (-0.73) \\ 
  Saturn Sun Venus & 62 & 6 & 0 & 10 & 16 & -4/16 (-0.25) \\ 
  Saturn Venus Mercury & 86 & 8 & 1 & 12 & 21 & -4/21 (-0.19) \\ 
  Saturn Venus Moon & 68 & 3 & 0 & 18 & 21 & -15/21 (-0.71) \\ 
  Jupiter Mars Sun & 51 & 7 & 0 & 8 & 15 & -1/15 (-0.07) \\ 
  Jupiter Mars Venus & 66 & 12 & 0 & 10 & 22 & 2/22 (0.09) \\ 
  Jupiter Mars Mercury & 68 & 19 & 1 & 3 & 23 & 16/23 (0.70) \\ 
  Jupiter Mars Moon & 58 & 16 & 0 & 5 & 21 & 11/21 (0.52) \\ 
  Jupiter Sun Venus & 79 & 17 & 1 & 5 & 23 & 12/23 (0.52) \\ 
  Jupiter Sun Mercury & 55 & 12 & 1 & 3 & 16 & 9/16 (0.56) \\ 
  Jupiter Sun Moon & 63 & 10 & 0 & 14 & 24 & -4/24 (-0.17) \\ 
  Jupiter Venus Mercury & 74 & 26 & 0 & 2 & 28 & 24/28 (0.86) \\ 
  Jupiter Venus Moon & 90 & 13 & 3 & 12 & 28 & 1/28 (0.04) \\ 
  Jupiter Mercury Moon & 69 & 31 & 2 & 0 & 33 & 31/33 (0.94) \\ 
  Mars Sun Venus & 41 & 6 & 0 & 10 & 16 & -4/16 (-0.25) \\ 
  Mars Sun Mercury & 78 & 6 & 0 & 11 & 17 & -5/17 (-0.29) \\ 
  Mars Sun Moon & 49 & 7 & 2 & 7 & 16 & 0/16 (0) \\ 
  Mars Venus Mercury & 55 & 10 & 2 & 11 & 23 & -1/23 (-0.04) \\ 
  Mars Venus Moon & 61 & 4 & 1 & 14 & 19 & -10/19 (-0.53) \\ 
  Mars Mercury Moon & 54 & 4 & 6 & 12 & 22 & -8/22 (-0.36) \\ 
  Sun Venus Mercury & 48 & 12 & 3 & 2 & 17 & 10/17 (0.59) \\ 
  Sun Venus Moon & 55 & 8 & 0 & 7 & 15 & 1/15 (0.07) \\ 
  Sun Mercury Moon & 42 & 15 & 1 & 1 & 17 & 14/17 (0.82) \\ 
  Venus Mercury Moon & 43 & 16 & 1 & 4 & 21 & 12/21 (0.57) \\ \hline
   \textbf{Total} & 2,145 & 337 & 28 & 270 & 635 & 66/635 (0.10)
\end{tabular}
\label{tab3}
\end{table}

\newpage 

\section*{Appendix}

\noindent Here, we consider the mathematical properties of random strings of digits (equivalent to random Z-codes) to illustrate the general combinatorial properties of these configurations. 

\subsection*{Expected number of conjunctions}

The number of words of length $n$ that use exactly $k$ letters from an alphabet of size $m$ is given by

\begin{equation}
    N(k, m, n) = {m \choose k} \cdot k! \cdot S_n^{(k)} = \frac{m!}{(m-k)!}\cdot S_n^{(k)}
\end{equation}

where $S_n^{(k)}$ is the Stirling number of the second kind. The total number of possible words is $m^n$. So, the probability of a randomly selected word using exactly $k$ letters is

\begin{equation}
    p_k^{m,n} = \frac{N(k, m,n)}{m^n} = \frac{m!}{(m-k)!}\cdot \frac{S_n^{(k)}}{m^n}
\end{equation}

For a Z-code we have $n$=7, $m$=12. So

\begin{equation}
    p_k=\frac{12!}{(12-k)!}\cdot\frac{S_7^{(k)}}{12^7}.
\end{equation}

When $k$=1, only one letter is used i.e. we have a septuple. When $k$=2, two letters are used so there must be a quintuple and a double, and so on. No repeated letters corresponds to $k$=7 (equivalent to no conjunctions), which is

\begin{equation}
\begin{split}
    p_{(k=7)} &= \frac{12!}{(12-7)!}\cdot \frac{S_7^{(7)}}{12^7} \\
    &= \frac{12!}{(12-7)!}\cdot  \frac{1}{12^7}\\
    &= \frac{385}{3456} \\
    &\approx 0.111
\end{split}
\end{equation}

The probability of at least one conjunction is given by 1-$p_{(k=7)}$. So $\approx$89\% of random Z-codes contain at least one conjunction. This is lower than the observed value of 97.5\%. For at least $x$ distinct conjunctions, the proportion is

\begin{equation}
    P(x)= \sum_{k=1}^x p_k 
\end{equation}

\begin{table}
\caption{Proportion of Z-codes containing minimum numbers of (separate) conjunctions. }
\label{tab_appendix}       
\centering
\begin{tabular}{l l l}
\hline\noalign{\smallskip}
Number of conjunctions & Random (\%)	& Observed (\%)\\\hline
0 & 11.1 &	2.5\\
1 or more &	88.9 &	97.5\\
2 or more &	49.9 &	42.7\\
3 or more &	12.7 &	8.4\\
4 or more &	1.1 &	0.9
\end{tabular}
\end{table}

Table \ref{tab_appendix} shows this proportion compared between real (0-200 CE) and random Z-codes.

\subsection*{Aspectual relationships}

\begin{figure*}
\centering
  \includegraphics[width=\textwidth]{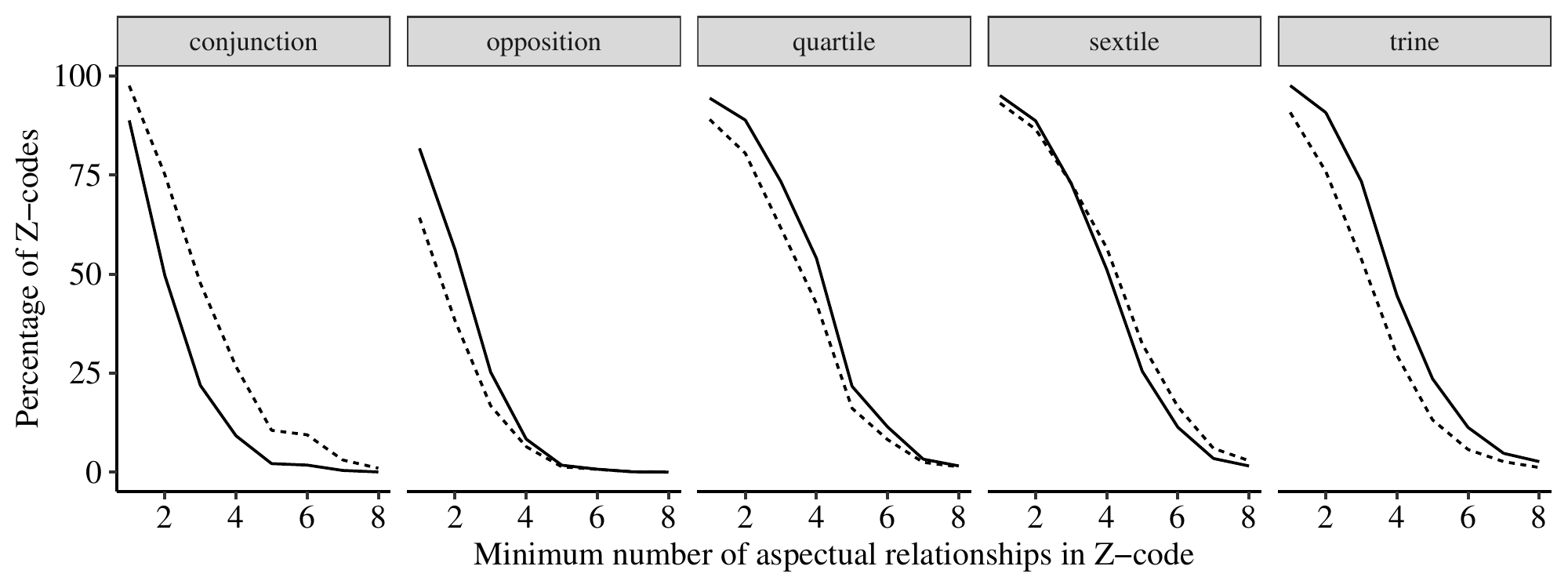}
\caption{The proportion of Z-codes which contain a minimum number of each (double) aspectual relationship. Random Z-codes (full line) and real Z-codes from 0-200 CE (dashed). For example, real Z-codes contain more conjunctions than random Z-codes, but in general the behaviour is similar.}
\label{fig_aspects}      
\end{figure*}

Figure \ref{fig_aspects} shows the minimum number of aspectual relationships in both real and random Z-codes, subsetted by the type of relationship. While there are some differences{---}such as a greater proportion of real Z-codes with conjunctions{---}the distributions are similar.

\subsection*{Minimum number of total aspectual relationships}

Conjunction is not the only relationship that the planets can have. In terms of the difference between the positions of two planets $a$ and $b$, we recall that the zodiac is a ring. So to get the minimum distance $D$ between their positions, we have 

\begin{equation}
    D(a, b) = \mathrm{min} \{ \lvert a-b \rvert, 12- \lvert a-b \rvert \}
\end{equation}

For example, in hexadecimal if $a=B$ (11) and $b=0$, then $|a-b|=B$ so $D(a,b)=1$. $D$ must be between 0 and 6. $D=0$ corresponds to conjunction. Other aspects are related to other values of $D$. The defined `aspects' of interest are 0 (conjunction), 2 (sextile), 3 (square), 4 (trine), 6 (opposition). 

So for a 7-digit string, we have $7 \choose 2$=21 combinations of planets and thus 21 possible values of $D$. We can separate these into two sorts:

\begin{itemize}
    \item \textit{aspectual} relationships: $D \in \{0, 2, 3, 4, 6\}$ 
    \item \textit{non-aspectual} relationships: $D \in \{1, 5\}$
\end{itemize} 

Let $I_\mathrm{aspect}$ be the number of aspectual relationships and $I_\mathrm{non-aspect}$ the number of non-aspectual relationships. We know that 

\begin{equation*}
    I_\mathrm{aspect}+I_\mathrm{non-aspect}=21. 
\end{equation*}

From inspecting Z-codes, we hypothesise that $I_\mathrm{aspect}^\mathrm{min}=9$. This is equivalent to the statement that 

\begin{equation*}
    I_\mathrm{non-aspect}^\mathrm{max}=21-9=12.
\end{equation*}

\noindent \textbf{Proof.} The proof rests on a useful fact about aspectual relationships.

\begin{quote}
\noindent    \textbf{Fact 1.} \textit{If planet $A$ is not in aspect with planets $B$ and $C$, then $B$ and $C$ are in aspect with each other.}
    
\noindent    To see that this is true, consider that all the four possible non-aspectual positions relative to planet $A$ (on the ring of the zodiac relative to $A$: 1, 5, 7, 11) are aspectual with respect to each other. Also any position is aspectual with respect to itself (0=conjunction). So, any placement of $B$ and $C$ in these non-aspectual positions (with respect to $A$) means they must be in aspect with each other.
\end{quote}

\noindent Now, there are two cases to consider. 

\vspace{2ex}\noindent \textbf{Case 1: one planet has at least 4 non-aspectual relationships}

\noindent We place $a$ and 4 planets $w, x, y, z$ such that $a$ has non-aspectual links with all of them. By \textbf{Fact 1}, there are no non-aspectual links between any pair of $w, x, y, z$.

Now add planet $e$. It can either have non-aspectual links with $a$ or with $w, x, y, z$, but not both (by \textbf{Fact 1}).

If $e$ has a non-aspectual link with $w, x, y, z$, then it has at most +4 non-aspectual links. Now considering a seventh planet, $f$. Similarly it can either have non-aspectual links with $a$ or with $w, x, y, z$ or with $e$, but not more than one of those. So at most +4 non-aspectual links. So $I_\mathrm{non-aspect}$ is at most $4+4+4=12$. 

Alternatively, if $e$ has a non-aspectual link with $a$ then this adds +1 non-aspectual link. (At this stage $I_\mathrm{non-aspect}=4+1=5$ and that means $f$ would need $>7$ non-aspectual links, which is not possible since there are only seven planets.) But for completeness, now add $f$. Similarly it can have at most links with $e$ and $w,x,y,z$. So at most +5 non-aspectual links. So $I_\mathrm{non-aspect}$ is at most $4+1+4=9$. 

\vspace{2ex} \noindent \textbf{Case 2: no planet has more than 3 non-aspectual links}

\noindent In this case, $I_\mathrm{non-aspect}$ cannot be larger than $\frac{3\times7}{2}=10.5$ which is less than 12.  

\vspace{2ex}

\noindent \textbf{Case 1} and \textbf{Case 2} cover all possible cases. So we have now shown that $I_\mathrm{non-aspect}^\mathrm{max}=12$. Therefore, we have shown that $I_\mathrm{aspect}^\mathrm{min}=9$. That is, any placement of seven planets in the twelve signs of the zodiac will produce at least nine aspectual relationships. So, every horoscope contains at least nine instances of double \textit{synkrasis} of planets.

\end{document}